\documentclass[showpacs,preprintnumbers,nofootinbib,
superscriptaddress,amsmath,floatfix,prd]{revtex4}

\usepackage{amssymb}  
\usepackage{graphicx}
\usepackage{exscale}
\usepackage{textcomp}
\usepackage{enumerate}

\usepackage[dvips]{color}

\usepackage[normalem]{ulem}  

\newcommand\ra{\rangle}
\newcommand\la{\langle}

\newcommand{\non}{\nonumber\\}

\newcommand{\be}{\begin{equation}}
\newcommand{\ee}{\end{equation}}
\newcommand{\bea}{\begin{eqnarray}}
\newcommand{\eea}{\end{eqnarray}}
\newcommand{\ba}[1]{\begin{array}{#1}}
\newcommand{\ea}{\end{array}}

\newcommand{\Tr}{{\rm Tr}}

\begin{document}

\title{Ginzburg-Landau phase diagram for dense matter with axial anomaly, strange quark mass, and meson condensation}

\author{Andreas Schmitt}
\email{aschmitt@hep.itp.tuwien.ac.at}
\affiliation{Institut f\"{u}r Theoretische Physik, Technische Universit\"{a}t Wien, 1040 Vienna, Austria}

\author{Stephan Stetina}
\email{stetina@hep.itp.tuwien.ac.at}
\affiliation{Institut f\"{u}r Theoretische Physik, Technische Universit\"{a}t Wien, 1040 Vienna, Austria}

\author{Motoi Tachibana}
\email{motoi@cc.saga-u.ac.jp}
\affiliation{Department of Physics, Saga University, Saga 840-8502, Japan}

\date{February 1, 2011}

\begin{abstract}

We discuss the phase structure of dense matter, in particular the nature of the transition between
hadronic and quark matter. Calculations within a Ginzburg-Landau approach show that the axial anomaly can induce a critical point in this
transition region. This is possible because in three-flavor quark matter with instanton effects a chiral condensate can be added to the 
color-flavor locked (CFL) phase
without changing the symmetries of the ground state. In (massless) two-flavor quark matter such a critical point is not possible since the 
corresponding color superconductor (2SC) does not break chiral symmetry. We study the effects of a nonzero
but finite strange quark mass which interpolates between these two cases.
Since at ultra-high density the first reaction of CFL to a nonzero strange quark mass is to develop a kaon condensate, we extend previous
Ginzburg-Landau studies by including such a condensate. We discuss the fate of the critical point  
systematically and show that the continuity between hadronic and quark matter can be disrupted by the onset of a kaon condensate. 
Moreover, we identify the mass terms in the Ginzburg-Landau potential which are needed for the 2SC phase to occur in the phase diagram.

\end{abstract}


\maketitle

\section{Introduction}
\label{intro}

Cold, strongly interacting  matter appears as ordinary nuclear matter at small baryon chemical potential and as 
color-flavor locked (CFL) quark matter \cite{Alford:1998mk,Alford:2007xm} at sufficiently large baryon chemical potential. 
The former is an observational fact while the latter
follows theoretically from first-principle calculations within perturbative QCD. Since QCD is asymptotically free \cite{Gross:1973id,Politzer:1973fx}
these calculations are valid at asymptotically large chemical potential. How matter evolves from nuclear matter to CFL quark matter,
i.e., which phases and phase transitions it passes in between, is an unsolved problem. This problem is of theoretical relevance since
it addresses a poorly understood territory of the QCD phase diagram. It is also of phenomenological relevance since the interior 
of compact stars contains such ``in-between'' matter, i.e., matter neither well described by low-density nuclear physics nor by ultra-high-density
perturbation theory (for an introduction to dense matter in compact stars see Ref.\ \cite{Schmitt:2010pn}). 

From the symmetries of ordinary nuclear (not hypernuclear) matter and the CFL phase it is clear that there must be at least one true phase 
transition between these two phases. However, since CFL breaks chiral symmetry it is very similar to hadronic matter and in principle 
allows for the intriguing possibility of a quark-hadron continuity \cite{Schafer:1998ef}. In other words, if hadronic matter is ``prepared'' 
by a series of 
appropriate phase transitions, including onsets of hyperonic matter and subsequent transitions to hyperon superfluids, 
it can become indistinguishable from CFL quark matter \cite{Alford:1999pa}. Whether these transitions happen, in particular whether 
they happen before deconfinement sets in, depends on the value of the constituent strange quark mass and the details of the baryonic interactions. 

We may also think of the quark-hadron transition  
by starting from the high-density limit. Upon going down in density, the effect of the strange quark mass becomes more important, imposing
a stress on the symmetric pairing pattern of CFL. Systematic studies at high densities show that CFL reacts on this stress first by
developing a kaon condensate \cite{BedaqueSchaefer}, followed by an anisotropic phase with a kaon supercurrent 
\cite{Schafer:2005ym,Kryjevski:2008zz,Schmitt:2008gn}, and then possibly a crystalline phase \cite{Alford:2000ze,Mannarelli:2006fy,Rajagopal:2006ig}. Also pairing patterns which do not break chiral symmetry may arise, such as the 2SC phase \cite{Bailin:1979nh} or spin-one color
superconductors \cite{Schafer:2000tw,Schmitt:2002sc,Schmitt:2004et}. Only if such a phase does not appear before confinement sets in, a smooth 
quark-hadron crossover is possible. 

The problem of the quark-hadron continuity can thus be thought of in the following illustrative way. 
Start from low (high) density and go up (down) in density and determine all phases on the way, say at zero temperature and 
for the moment ignoring the deconfinement (confinement) transition.
Although we currently do not have the theoretical tools to determine all these phases in detail, this will in principle 
lead to two sequences of 
phases. One describes strongly interacting hadronic matter between ordinary nuclear matter at low density and some phase at very high density, 
the other describes quark matter between CFL at ultra-high density and some phase at lower density. Now we would like to put these two 
sequences together to obtain the true QCD phase diagram at small temperatures. We know that they fit 
smoothly if we glue them together at the point of the CFL phase in the quark matter part and the point of the hyperon superfluid phase described in 
Ref.\ \cite{Alford:1999pa} in the hadronic part. 

In this paper we address this problem in a very simplified fashion. Our first simplification is the use of a Ginzburg-Landau free energy as in 
Refs.\ \cite{Hatsuda:2006ps,Yamamoto:2007ah}, see also Refs.\ \cite{Iida:2000ha,Iida:2001pg,Iida:2002ev,Iida:2003cc,Giannakis:2004xt,Iida:2004cj};
for recent discussions of the quark-hadron transition in a Nambu-Jona-Lasinio (NJL) model see Refs.\ \cite{Abuki:2010jq,Basler:2010xy,Wang:2010iu}. 
Within the Ginzburg-Landau approach
we will not be able to make firm predictions for QCD. The benefit is, however, that our study is model-independent and systematic since
the free energy is based on the underlying symmetries and we can in principle determine and classify all phases in the parameter space of the 
Ginzburg-Landau coefficients. The order parameters we use are the chiral condensate and the diquark condensate. Our ``nuclear matter'' is thus
mimicked simply by a nonvanishing chiral condensate, and we cannot account for the various phase transitions which potentially 
``prepare'' hadronic matter for the quark-hadron crossover. The onset of a CFL diquark condensate is a very crude version of these transitions. 
After this onset has happened, the question
is whether the chiral condensate can approach zero smoothly, i.e., without causing an additional phase transition. For the case
of a vanishing strange quark mass this is possible, and the Ginzburg-Landau parameter space where this is realized has been determined in 
Refs.\ \cite{Hatsuda:2006ps,Yamamoto:2007ah}. In particular, it can only be realized in the presence of a term that violates the $U(1)_A$
symmetry and thus is an effect of the axial anomaly (other effects of the anomaly in high-density QCD are discussed for instance in 
Refs.\ \cite{Rapp:1999qa,Schafer:2002ty}). Only then do the CFL phase without and the CFL phase with chiral condensate 
have the same residual symmetry. As a consequence, the $U(1)_A$-violating term -- which couples chiral and diquark condensates -- 
induces a critical point in the Ginzburg-Landau phase diagram where a first-order phase transition line ends. 
The Ginzburg-Landau study also shows that for infinite strange quark mass, i.e., in two-flavor quark matter,
there is always a true phase transition from the phase where both chiral and diquark condensates are nonzero to the phase where only
the diquark condensate is nonvanishing. 

We know that the real world is somewhere in between, i.e., the strange quark mass is neither negligibly small nor approximately infinite
in the region of intermediate densities we are interested in. Therefore, in this paper we extend the previous Ginzburg-Landau studies by 
including a nonzero but finite strange quark mass. The first obvious consequence is the appearance of additional terms in the Ginzburg-Landau
free energy. Additionally, we take into account meson condensation in CFL. Since CFL breaks chiral symmetry, there is an octet of Goldstone modes, 
just as in the hadronic phase. At high density an effective theory for these modes, based on the symmetries of CFL, 
can be derived \cite{BedaqueSchaefer}, and it can be shown that the masses of the Goldstone modes are inversely ordered compared to the ordinary 
mesons in hadronic matter \cite{Son:1999cm}. The high-density effective theory
suggests that the neutral kaons form a Bose condensate in the presence of a nonzero strange quark mass \cite{BedaqueSchaefer}. The
resulting phase has been termed CFL-$K^0$ (see 
for instance Refs. \cite{Kaplan:2001qk,Alford:2007qa,Alford:2008pb} for properties of CFL-$K^0$ computed from the high-density
effective theory and Refs.\ \cite{Buballa:2004sx,Forbes:2004ww,Warringa:2006dk,Basler:2009vk} for calculations within an NJL model). 
It is thus natural to include kaon condensation into the Ginzburg-Landau calculation. We note, however, that at lower densities
the meson masses may receive large instanton corrections \cite{Schafer:2002ty}. 
More formally speaking, at ultra-high densities, where instanton
effects are suppressed, the meson masses squared are quadratic in the quark masses since a linear term in the masses is forbidden
by the $U(1)_A$ symmetry. When a linear, $U(1)_A$-violating term is allowed at lower densities, the masses -- and thus also their ordering --
is subject to potentially large corrections and it is unclear which of the mesons form a Bose condensate. Therefore, our
choice of a kaon condensate is inspired by the high-density results, but in principle also other meson condensates should be considered. 

Our paper is organized as follows. In Sec.\ \ref{sec:GL} we set up the Ginzburg-Landau potential, including mass terms and 
kaon condensate. Then we first evaluate the potential in the limit case of vanishing kaon condensate. This is done in Sec.\ \ref{sec:ms}. 
The purpose of this section is to 
explain the Ginzburg-Landau phase diagram in terms of the symmetries of the various phases and in the presence of a mass-induced
linear term in the chiral potential. We build on the phase diagrams obtained in this section when we include the effect of a kaon  
condensate in Sec.\ \ref{sec:K0}. In Sec.\ \ref{sec:2SC} we discuss the appearance of the 2SC phase due to mass corrections 
in the Ginzburg-Landau potential before we give our conclusions in Sec.\ \ref{sec:conclusions}.

\section{Ginzburg-Landau potential}
\label{sec:GL}

We are interested in a Ginzburg-Landau potential for the chiral and diquark condensates, including a neutral kaon condensate and 
corrections from the strange quark mass, based on Refs.\ \cite{Hatsuda:2006ps,Yamamoto:2007ah}.
The final result of this section is Eq.\ (\ref{Omfull}), and the following pages are devoted to the derivation of this equation.

\subsection{Symmetries and order parameters}

We shall consider a Ginzburg-Landau free energy of the form 
\be
\Omega = \Omega_\Phi + \Omega_d + \Omega_{\Phi d} \, ,
\ee
with a chiral part $\Omega_\Phi$ depending only on the chiral condensate $\Phi$, a diquark part depending only on the diquark condensates
$d_L$ and $d_R$ (which in turn depend on the kaon condensate) and an interaction part $\Omega_{\Phi d}$ which couples $\Phi$ with $d_L$ and $d_R$.
In terms of left- and right-handed quark fields $q_L$ and $q_R$ we have $\la \bar{q}_{Ri}^\alpha q_{Lj}^\alpha \ra \propto \Phi_{ji}$
and $\la q_{Li}^\alpha C q_{Lj}^\beta \ra \propto -\epsilon^{\alpha\beta A}\epsilon_{ijB}
[d_L^\dag]_B^A  $, $\la q_{Ri}^\alpha C q_{Rj}^\beta \ra \propto \epsilon^{\alpha\beta A}\epsilon_{ijB}
[d_R^\dag]_B^A  $, with flavor indices $i,j,B$, color indices $\alpha,\beta, A$, and the charge conjugation matrix $C=i\gamma^2\gamma^0$.
Since we consider a three-flavor system, $\Phi$, $d_L$, and $d_R$ are $3\times 3$ matrices.
The quark fields transform under the symmetry group
\be \label{G}
G \equiv SU(3)_c \times SU(3)_L \times SU(3)_R \times U(1)_B \times U(1)_A
\ee
as 
\be \label{qtrafo}
q_L \to e^{i\alpha_B}e^{-i\alpha_A} V_L U \,q_L \, , \qquad q_R \to e^{i\alpha_B}e^{i\alpha_A} V_R U \,q_R \, ,
\ee
where $(V_L,V_R)\in SU(3)_L \times SU(3)_R$ is a chiral transformation, $U\in SU(3)_c$ is a color gauge transformation,  
\mbox{$e^{i\alpha_B}\in U(1)_B$} 
is a transformation associated with baryon number conservation, and $e^{i\alpha_A}\in U(1)_A$ is an axial transformation. 
In terms of left- and right-handed $U(1)$ transformations we have $q_L\to e^{i\alpha_L}q_L$, $q_R\to e^{i\alpha_R}q_R$, hence the vector
and axial $U(1)$ transformations in Eq.\ (\ref{qtrafo}) follow from $\alpha_B = (\alpha_R+\alpha_L)/2$, $\alpha_A = (\alpha_R-\alpha_L)/2$.
Eventually, our potential will not be invariant under the full group $G$. The chiral group $SU(3)_L \times SU(3)_R$ and the axial $U(1)_A$
become approximate symmetries after including the effects of a small strange quark mass and the QCD axial anomaly, respectively.

The transformation properties of the order parameters under $G$ are
\be \label{Phitrafo}
\Phi \to e^{-2i\alpha_A} V_L\Phi V_R^\dag \, , 
\ee  
and 
\be \label{dLdRtrafo}
d_L \to e^{-2i\alpha_B}e^{2i\alpha_A} V_L d_L U^T \, , \qquad 
d_R \to e^{-2i\alpha_B}e^{-2i\alpha_A} V_R d_R U^T \, .
\ee
The mass terms are generated by the field 
\be
M = \left(\begin{array}{ccc} m_u&0&0\\ 0&m_d&0 \\ 0&0&m_s\end{array}\right) \, ,
\ee
which transforms under $G$ in the same way as the chiral field $\Phi$,
\be
M \to e^{-2i\alpha_A} V_L M\, V_R^\dag \, . 
\ee 
Although we shall write down the Ginzburg-Landau terms with general quark masses $m_u$, $m_d$, $m_s$, we shall later neglect the 
up and down quark masses and only keep the strange quark mass.  

Our ansatz for the order parameters is as follows. The chiral field is  
\be
\Phi = \left(\begin{array}{ccc} \sigma_u&0&0\\ 0&\sigma_d&0 \\ 0&0&\sigma_s\end{array}\right) \, .
\ee
We shall derive the potential $\Omega$ within this general ansatz, but later set for simplicity $\sigma_u=\sigma_d=\sigma_s$. Different values
for each quark flavor are more realistic in the presence of a strange quark mass and a kaon condensate. However, this would introduce
additional independent parameters into our potential, making a systematic evaluation very complicated. Therefore, we shall use the symmetric 
case as a simplification. 

For the diquark condensate we use the ansatz
\be \label{dLdR}
d_L=d_R^\dagger = d\left(\begin{array}{ccc} 1&0&0\\ 0&\cos(\phi/2)&i\sin(\phi/2) \\ 0&i\sin(\phi/2)&\cos(\phi/2)\end{array}
\right) \, , 
\ee
where $\phi$ is the kaon condensate. For $\phi=0$ we recover the pure CFL order parameter
$d_L=d_R={\rm diag}(d,d,d)$. A nonzero $\phi$ introduces a relative rotation between left- and right-handed diquarks. 
We have chosen a neutral kaon condensate which is the most likely possibility at high 
densities but, as explained in the introduction, at intermediate densities there could also be other meson condensates. 

As a comparison, note that 
the chiral field in the high-density effective theory of CFL \cite{BedaqueSchaefer} is 
\be
\Sigma = e^{i\theta_a\lambda_a/f_\pi} \, ,
\ee
with the meson fields $\theta_a$, the Gell-Mann matrices $\lambda_a$, and the (CFL version of the) pion decay constant $f_\pi$.
Ignoring all meson fields other than the neutral kaon, we set all $\theta_a$'s to zero except for $\theta_6$ and $\theta_7$. Without
loss of generality we can also set $\theta_7=0$. Then, our condensate $\phi$ is the vacuum expectation value of $\theta_6/f_\pi$, and 
the chiral field is written in terms of the diquark condensates as 
\be \label{Sigma}
\Sigma = \frac{d_Ld_R^\dagger}{d^2} = 
\left(\begin{array}{ccc} 1&0&0\\ 0&\cos \phi &i\sin \phi  \\ 0&i\sin \phi &\cos \phi \end{array}
\right)\, . 
\ee
While $d_L$ and $d_R$ are gauge variant quantities, $\Sigma$ is gauge invariant. Therefore, our ansatz (\ref{dLdR}) is one of infinitely many 
choices for $d_L$ and $d_R$ -- all related by gauge transformations -- which lead to Eq.\ (\ref{Sigma}). The chiral field
transforms under $G$ as  
\be
d_Ld_R^\dagger \to e^{4i\alpha_A} V_L d_Ld_R^\dagger V_R^\dag \, .
\ee
This is the same transformation property as the ordinary chiral field $\Phi$, except for the transformations under $U(1)_A$. 
This difference reflects the fact that in CFL the mesons are composed of four quarks, not two.

\subsection{Chiral potential}

We can now derive the explicit form of the potential $\Omega$. For the chiral part $\Omega_\Phi$ we collect all terms up to fourth combined order 
in $M$ and $\Phi$ with at most one power in the mass field $M$. The terms of ${\cal O}(M^0)$ which are invariant 
under $G$ are
\begin{subequations} \label{phionly}
\bea
\Tr[\Phi^\dag\Phi] &=& \sigma_u^2+\sigma_d^2+\sigma_s^2 \, ,\\
(\Tr[\Phi^\dag\Phi])^2 &=& (\sigma_u^2+\sigma_d^2+\sigma_s^2)^2 \, ,\\
\Tr[(\Phi^\dag\Phi)^2] &=& \sigma_u^4+\sigma_d^4+\sigma_s^4 \, . 
\eea
\end{subequations}
If we do not require the potential to be invariant under $U(1)_A$, we have the additional term \cite{Pisarski:1983ms}
\be \label{phianom}
{\rm det} \,\Phi + {\rm h.c.} \propto \epsilon_{abc}\epsilon_{ijk}\Phi_{ai}\Phi_{bj}\Phi_{ck} + {\rm h.c.}  
= 12\sigma_u\sigma_d\sigma_s \, , 
\ee
which, because of  
\be \label{anomtrafo}
{\rm det}\,\Phi \to e^{-6i\alpha_A} {\rm det}\,\Phi \, , 
\ee
is only invariant under the discrete subgroup $\mathbb{Z}_A(6)\subset U(1)_A$, as expected from the anomaly. Microscopically, 
${\rm det}\,\Phi$ accounts for an effective six-point instanton vertex which converts three left-handed quarks into
three right-handed quarks, thus violating axial charge conservation by an amount $2N_f = 6$. 

The terms of order ${\cal O}(M^1)$ arise from replacing one chiral field $\Phi$ by the mass field $M$ in each of the above ${\cal O}(M^0)$ terms.
We obtain
\begin{subequations} \label{phiM}
\bea
\Tr[M\Phi^\dag]+ {\rm h.c.} &=& 2(m_u\sigma_u+m_d\sigma_d+m_s\sigma_s) \, , \label{phiM1} \\
\Tr[M\Phi^\dag]\Tr[\Phi^\dag\Phi] + {\rm h.c.} &=& 2(m_u\sigma_u+m_d\sigma_d+m_s\sigma_s)(\sigma_u^2+\sigma_d^2+\sigma_s^2) \, , \\
\Tr[M\Phi\Phi^\dag\Phi] + {\rm h.c.} &=& 2(m_u\sigma_u^3+m_d\sigma_d^3+m_s\sigma_s^3) \, , 
\eea
\end{subequations}
and the anomalous term becomes
\be \label{phiManom}
\epsilon_{abc}\epsilon_{ijk}M_{ai}\Phi_{bj}\Phi_{ck} + {\rm h.c.} = 4(\sigma_u\sigma_d m_s + \sigma_u\sigma_s m_d + \sigma_d\sigma_s m_u) \, . 
\ee
We can now add the contributions (\ref{phionly}), (\ref{phianom}), (\ref{phiM}), (\ref{phiManom}), each coming with a separate 
prefactor, to obtain the chiral potential. Approximating $m_u\simeq m_d\simeq 0$ and setting for simplicity 
$\sigma_u = \sigma_d = \sigma_s \equiv \sigma$, we can write the potential as 
\bea \label{OmPhi}
\Omega_\Phi &=& a_0 m_s\sigma + \frac{a_1+m_s a_2}{2}\sigma^2+\frac{c_1+m_sc_2}{3}\sigma^3 +\frac{b}{4}\sigma^4 \, .
\eea
In the given approximation, $m_s$ 
gives rise to a linear term in $\sigma$ and yields corrections to the quadratic and cubic terms. 
Because of the linear term, the chiral condensate cannot vanish exactly in the ground state. Instead of a vacuum phase with $\sigma=0$ there
will be a phase with very small $\sigma$ where chiral symmetry is approximately restored and which is continuously connected to the 
chirally broken phase in which $\sigma$ has a sizable value. This is the most obvious consequence of the mass term. Due to the coupling
of chiral and diquark condensates, to be discussed in Sec.\ \ref{sec:interaction}, we shall find other, less obvious, effects of the linear term 
for our phase diagram. These effects are discussed in Sec.\ \ref{sec:ms}.

\subsection{Diquark potential and comparison to high-density effective theory}
\label{sec:diquark}

For the diquark potential $\Omega_d$ we also start from the terms up to ${\cal O}(d^4)$, first without mass insertions. Within our 
CFL-$K^0$ ansatz they simply yield the structures $d^2$ and $d^4$ since the kaon condensate always drops out,
\begin{subequations}\label{dtraces}
\bea
\Tr[d_Ld_L^\dag] &=&\Tr[d_Rd_R^\dag] = 3d^2 \, , \\
(\Tr[d_Ld_L^\dag])^2 &=&(\Tr[d_Rd_R^\dag])^2 = \Tr[d_Ld_L^\dag]\Tr[d_Rd_R^\dag] \non 
=3\Tr[(d_Ld_L^\dag)^2] &=& 3\Tr[(d_Rd_R^\dag)^2] = 3\Tr[d_Rd_L^\dag d_Ld_R^\dag] = 
 9d^4 \, . 
\eea
\end{subequations}
All these terms are invariant under the full group $G$. There is no such term as ${\rm det} \, d_{L,R}$ since this term would not only
break $U(1)_A$ but also baryon number conservation which must not be explicitly broken.  
To include the effect of quark masses, we need to replace $d_Ld_R^\dag$ by $M$. 
From Eqs.\ (\ref{dtraces}) the only possible term (except for a term constant in $d_L$, $d_R$ which we can omit) is 
\be \label{Mdd}
\Tr[d_L^\dag Md_R]+{\rm h.c.} = 2d^2[m_u+(m_d+m_s)\cos\phi] \, .
\ee
This term is invariant under $\mathbb{Z}_A(6) \subset U(1)_A$ and thus is allowed in the presence of the anomaly. 
Other ${\cal O}(M^1)$ terms arise from replacing one chiral field $d_Ld_R^\dag$ in the ${\cal O}(d^6)$ terms. They all yield the same structure
and are invariant under $\mathbb{Z}_A(6)$,
\bea \label{md4}
&& 3(\Tr[d_L^\dag Md_R d_R^\dag d_R ] +{\rm h.c.}) = 3(\Tr[d_L^\dag Md_R d_L^\dag d_L ] +{\rm h.c.}) \non
&&= \Tr[d_L^\dag Md_R]\Tr[d_R^\dag d_R] + {\rm h.c.}=\Tr[d_L^\dag Md_R]\Tr[d_L^\dag d_L] + {\rm h.c.}= 6d^4\left[m_u+(m_d+m_s)\cos\phi\right] \, .
\eea
So far the only structure we have produced for the kaon condensate is $\cos\phi$. If we were to stop here, the minimization of $\Omega_d$ would 
not allow for nontrivial condensates. This would be in contradiction to high-density calculations. Therefore we need to include at least 
one extra term with nontrivial structure in $\phi$. To find this term it is useful to briefly 
discuss the Lagrangian of the high-density effective theory,
\be \label{Leff}
{\cal L}_{\rm eff} = \frac{f_\pi^2}{4}\Tr[\nabla_0\Sigma\nabla_0\Sigma^\dag - v_\pi^2\partial_i\Sigma\partial_i\Sigma^\dag]
+B\Tr[M\Sigma^\dag + M^\dag\Sigma] + \frac{af_\pi^2}{2}{\rm det}M\,\Tr[M^{-1}(\Sigma+\Sigma^\dag)] \, , 
\ee
where weak-coupling calculations give the following values for the constants \cite{Son:1999cm,BedaqueSchaefer,Schafer:2002ty},
\be \label{fvaB}
f_\pi^2 = \frac{21-8\ln 2}{18}\frac{\mu_q^2}{2\pi^2} \, , \qquad v_\pi^2 = \frac{1}{3} \, , \qquad a = \frac{3\Delta^2}{\pi^2f_\pi^2}
\, , \qquad B = c
 \left(\frac{3\sqrt{2}\pi}{g}\Delta\,
     \frac{\mu_q^2}{2\pi^2}\right)^2
  \left(\frac{8\pi^2}{g^2}\right)^{6}
  \frac{\Lambda_{\rm QCD}^9}{\mu_q^{12}} \, .
\ee
Here, $\mu_q$ is the quark chemical potential, $\Delta$ the superconducting gap parameter, $c\simeq 0.155$, $g$ the strong coupling constant, and 
$\Lambda_{\rm QCD}$ the QCD scale factor. The term linear in $M$ with prefactor $B$ corresponds to the term in Eq.\ (\ref{Mdd}) and breaks $U(1)_A$. 
An important contribution to the effective Lagrangian is given by promoting the time derivatives to covariant derivatives,
\be \label{covariant}
\nabla_0\Sigma\equiv \partial_0\Sigma+i[A,\Sigma] \, , \qquad A\equiv -\frac{MM^\dag}{2\mu_q} \, .
\ee
Including only neutral kaons, i.e., using the chiral field from Eq.\ (\ref{Sigma}),
the tree-level potential $V(\phi) = -{\cal L}_{\rm eff}$ becomes [after subtracting the ``vacuum'' contribution $V(\phi=0)$]
\be \label{Vphi}
V(\phi) = -\frac{f_\pi^2}{2}\mu^2\sin^2\phi + f_\pi^2 m_{K^0}^2 (1-\cos\phi) \, ,   
\ee
with the effective kaon mass squared
\be \label{mK0}
m_{K^0}^2\equiv \frac{2B(m_d+m_s)}{f_\pi^2}+am_u(m_s+m_d) \, .
\ee
The kaon mass squared receives an instanton contribution linear in the quark masses 
\cite{Manuel:2000wm,Schafer:2002ty} and a contribution quadratic in the quark masses. In the weak-coupling limit at very high densities,
$\mu_q\gg \Lambda_{\rm QCD}$, the instanton contribution is negligible and $U(1)_A$ is effectively restored. In this case
the term $M^{-1}{\rm det}\, M$ becomes dominant, resulting in the inverse meson mass ordering \cite{Son:1999cm}. 
We do not include a term of ${\cal O}(M^2)$ in our 
Ginzburg-Landau potential, but nevertheless assume the kaon to be the most likely meson to condense.

The covariant derivative  (\ref{covariant}) gives rise to the term $\propto \sin^2\phi$ which contains the effective kaon chemical 
potential 
\be \label{muK0}
\mu \equiv \frac{m_s^2-m_d^2}{2\mu_q} \, .
\ee
This term is crucial for kaon condensation to occur. One can see this for instance by expanding $V(\phi)$ for small $\phi$, say up to 
order $\phi^4$ to reproduce the potential of an ordinary $\phi^4$ model. The resulting $\phi^2$ term becomes negative only for $\mu>m_{K^0}$
in which case a nonzero value of $\phi$ minimizes the potential. We thus conclude that the critical strange quark mass 
for the onset of kaon condensation scales at high density as $m_s \sim m_u^{1/3}\Delta^{2/3}$. 

From Eq.\ (\ref{covariant}) we see that the term $\propto\mu^2$ in the potential is formally of order $M^4$. Nevertheless, it can become 
comparable to the $M^2$ terms \cite{BedaqueSchaefer}, and therefore we need to include this term in our 
potential,\footnote{Alternatively, we may write the diquark field as 
\begin{equation*} 
d_L=d_R^\dagger = d\left(\begin{array}{ccc} 1&0&0\\ 0&\cos(\phi/2)&ie^{i\mu t}\sin(\phi/2) \\ 0&ie^{-i\mu t}\sin(\phi/2)&\cos(\phi/2)\end{array}
\right) \, , 
\end{equation*}
including the effective kaon chemical potential $\mu$. Then, the time derivative term,
\begin{equation*} \label{dt}
-\Tr[\partial_0(d_Ld_R^\dag)\partial_0(d_Rd_L^\dag)]  = -2\mu^2d^4\sin^2\phi \, ,
\end{equation*}
produces the structure of Eq.\ (\ref{muterm}).
This way of introducing a meson chemical potential in the time dependence of the condensate is done for instance in the context of 
meson condensation in nuclear matter \cite{Baym:1973zk,Muto:1992pq,Thorsson:1993bu}.} 
\be \label{muterm}
\Tr[[MM^\dag,d_Ld_R^\dag][M^\dag M,d_Rd_L^\dag]] = -2d^4(m_d^2-m_s^2)^2\sin^2\phi \, .
\ee
Strictly speaking, in the spirit of the Ginzburg-Landau
expansion it is not consistent to keep one ${\cal O}(M^4)$ term and drop all ${\cal O}(M^2)$ terms. Guided by the high-density 
results we need to assume that the particular Ginzburg-Landau coefficient in front of the term (\ref{muterm}) is large enough to 
compensate for the suppression due to the quark mass. 

We can now collect all terms in Eqs.\ (\ref{dtraces}), (\ref{Mdd}), (\ref{md4}), (\ref{muterm}) and set $m_u\simeq m_d\simeq 0$ 
to write the diquark potential as
\bea \label{Omd}
\Omega_d &=& \frac{\alpha_1+\alpha_2m_s\cos\phi}{2}\,d^2 + \frac{\beta_1+\beta_2m_s\cos\phi-\mu^2\sin^2\phi}{4}\,d^4 \, ,
\eea
where we have written the prefactor of the term (\ref{muterm}) as $\mu^2$, reminiscent of the effective kaon chemical potential (\ref{muK0})
of the effective theory. 

\subsection{Interaction potential}
\label{sec:interaction}

For the interaction potential $\Omega_{\Phi d}$, the term of lowest order in the order parameters is 
\be \label{ddPhi}
\Tr[d_Rd_L^\dag\Phi]+{\rm h.c.} = 2d^2[\sigma_u+(\sigma_d+\sigma_s)\cos\phi] \, .
\ee
This term is anomalous since it is not invariant under $U(1)_A$. Without kaon condensate it has already been considered in 
Ref.\ \cite{Hatsuda:2006ps,Yamamoto:2007ah}. We see that, in contrast to the pure diquark terms, the kaon condensate appears even in the 
absence of a strange quark mass. Including one more power in $\Phi$ yields the following terms which are invariant under the full
symmetry group $G$,
\begin{subequations} 
\bea
\Tr[d_Ld_L^\dag  + d_Rd_R^\dag]\Tr[\Phi^\dag\Phi] &=& 6\Tr[d_Ld_L^\dag\Phi\Phi^\dag] = 6\Tr[d_Rd_R^\dag\Phi\Phi^\dag] = 
6d^2(\sigma_u^2+\sigma_d^2+\sigma_s^2) \, , \\
{\rm det}\Phi \,\Tr[d_Ld_R^\dag \Phi^{-1}]+{\rm h.c.} &=& 2d^2[\sigma_d\sigma_s + \sigma_u(\sigma_d+\sigma_s)\cos\phi] \, .
\eea
\end{subequations}
We include mass terms up to ${\cal O}(M^1)$ which arise from replacing a chiral field $\Phi$ in the above expressions (obviously,
replacing a field $d_Ld_R^\dag$ in these terms does not yield interaction terms),
\begin{subequations} \label{ddMPhi}
\bea
\Tr[d_Ld_L^\dag  + d_Rd_R^\dag]\Tr[M^\dag\Phi] +{\rm h.c.} &=& 6(\Tr[d_Ld_L^\dag M\Phi^\dag]+{\rm h.c.}) = 
12d^2(m_u\sigma_u+m_d\sigma_d+m_s\sigma_s) \, , \\
\epsilon_{abc}\epsilon_{ijk}M_{ai}\Phi_{bj}(d_Ld_R^\dag)_{ck}+{\rm h.c.} &=& 
2d^2\{m_d\sigma_s+m_s\sigma_d+[m_u(\sigma_d+\sigma_s)+\sigma_u(m_d+m_s)]\cos\phi\}
\, .
\eea
\end{subequations}
Estimates of the prefactor of the $d^2\sigma^2$ term suggest this contribution to be negligible \cite{Hatsuda:2006ps,Yamamoto:2007ah}. 
We shall thus focus only on the $d^2\sigma$ terms, as it was done for the three-flavor case in Refs.\ \cite{Hatsuda:2006ps,Yamamoto:2007ah}.
Then, collecting the terms in Eqs.\ (\ref{ddPhi}), (\ref{ddMPhi}), and again setting $m_u\simeq m_d\simeq 0$, 
$\sigma_u = \sigma_d = \sigma_s \equiv \sigma$, the most general form of the interaction potential can be written as 
\be \label{OmPhid}
\Omega_{\Phi d} =  -\left[\gamma\frac{1+2\cos\phi}{3} + m_s(\gamma_1 + \gamma_2\cos\phi)\right] d^2\sigma \, .
\ee
For $\phi=m_s=0$ we recover the interaction term $-\gamma d^2\sigma$ from Refs.\ \cite{Hatsuda:2006ps,Yamamoto:2007ah}. 

\subsection{Full potential}

We can now put the contributions (\ref{OmPhi}), (\ref{Omd}), (\ref{OmPhid}) together to obtain the full Ginzburg-Landau potential, including 
mass corrections and the meson condensate in the approximations discussed above,  
\bea \label{Omfull}
\Omega &=& a_0 m_s\sigma + \frac{a_1+m_s a_2}{2}\sigma^2+\frac{c_1+m_sc_2}{3}\sigma^3 +\frac{b}{4}\sigma^4 \non
&&+ \frac{\alpha_1+\alpha_2m_s\cos\phi}{2}\,d^2 + \frac{\beta_1+\beta_2m_s\cos\phi-\mu^2\sin^2\phi}{4}\,d^4  \non
&&-\left[\gamma\frac{1+2\cos\phi}{3} + m_s(\gamma_1 + \gamma_2\cos\phi)\right] d^2\sigma \, .
\eea
In the remainder of this paper we discuss and evaluate this potential, i.e., we shall solve the stationarity equations for 
the three order parameters $\sigma$, $d$, and $\phi$,
\be \label{stat}
\frac{\partial\Omega}{\partial \sigma} = \frac{\partial\Omega}{\partial d} = \frac{\partial\Omega}{\partial \phi} = 0 \, , 
\ee
in order to determine the ground state in the given parameter space. As written in Eq.\ (\ref{Omfull}) there are 14 independent parameters.
This is too unwieldy for a systematic study and therefore we shall work in several limit cases in the subsequent sections. 
In the next section we start with the case of a vanishing kaon condensate. 

\section{Mass effect without meson condensate}
\label{sec:ms}

For vanishing kaon condensate, $\phi=0$, all mass terms except for the linear $\sigma$ term are simply numerical corrections to the Ginzburg-Landau 
parameters. We thus absorb these mass terms into new overall prefactors. This reduces the number of independent parameters and 
does not change the results qualitatively (we do not attempt do determine the complete quantitative effect of the strange quark mass in the 
full parameter space). Then we can write the potential (\ref{Omfull}) as 
\bea \label{Om0}
\Omega(\sigma,d) &=& a_0\sigma + \frac{a}{2}\sigma^2-\frac{c}{3}\sigma^3 +\frac{b}{4}\sigma^4 + \frac{\alpha}{2} d^2 + \frac{\beta}{4} d^4  
-\gamma d^2\sigma \, .
\eea
The minus signs in front of the cubic term $\sigma^3$ and the interaction term $d^2\sigma$ are conventions which imply 
$c,\gamma>0$ \cite{Yamamoto:2007ah}. We also assume $b,\beta>0$ which ensures the boundedness of the free energy. The other 
parameters $a_0$, $a$, $\alpha$ shall be varied without further constraints. 
For notational convenience we have absorbed the factor $m_s$ into the definition of $a_0$. 
Without this linear term, the potential (\ref{Om0}) has been discussed in Refs.\ \cite{Hatsuda:2006ps,Yamamoto:2007ah}, where the phase structure
in the $(a,\alpha)$-plane has been given. We shall discuss the phase structure in the three-dimensional parameter space $(a_0,a,\alpha)$ 
and present cuts through this space for fixed values of $a_0$, such that we reproduce the known results for the special case $a_0=0$. 
We shall see that several analytical arguments used in the 
massless case regarding phase transitions and critical points can be used repeatedly also for the more complicated cases. For the complete 
evaluation of the phase diagram, however, we need to employ numerical calculations.

Without kaon condensate there are two stationarity equations,
\begin{subequations} \label{stat1}
\bea
0 &=& \frac{\partial\Omega}{\partial \sigma} = a_0 + a\sigma - c\sigma^2 + b\sigma^3 - \gamma d^2 \, , \label{stat1a}\\
0 &=& \frac{\partial\Omega}{\partial d} = \alpha d + \beta d^3 - 2\gamma d \sigma \, . \label{stat1b}
\eea
\end{subequations}
For the determination of the ground state it is useful to also consider the second derivatives. The Hessian matrix of $\Omega(d,\sigma)$
is
\be \label{hessian}
H = \left(\begin{array}{cc} \displaystyle{\frac{\partial^2\Omega}{\partial\sigma^2}} &
\displaystyle{\frac{\partial^2\Omega}{\partial\sigma\partial d}} \\[2ex] \displaystyle{\frac{\partial^2\Omega}{\partial d\partial \sigma}} 
&\displaystyle{\frac{\partial^2\Omega}{\partial d^2}} \end{array}\right)
= \left(\begin{array}{cc} a-2c\sigma +3b\sigma^2 & -2\gamma d \\[1.5ex] -2\gamma d & \alpha + 3\beta d^2 -2\gamma\sigma 
\end{array}\right) \, .
\ee   
For a solution of Eqs.\ (\ref{stat1}) to be a local minimum, the Hessian, evaluated at this solution, must have positive eigenvalues. If the 
potential is bounded from below -- which is guaranteed by $b,\beta>0$ -- the solution of Eqs.\ (\ref{stat1}) with the lowest free energy 
yields the ground state, i.e., the global minimum, unambiguously. In other words, if we find a stationary point of the potential
with a negative eigenvalue of its Hessian, then we must also have found another stationary point with lower free energy. Since the potential is
bounded from below, the stationary point with lowest free energy must have a positive definite Hessian. Consequently, 
the second derivatives are strictly speaking not necessary for a stability check.
We shall see below, however, that they yield a useful alternative method to determine the phase transition lines, some of which can be 
computed in an relatively simple way by using the eigenvalues of the Hessian.

Employing the terminology of Refs.\ \cite{Hatsuda:2006ps,Yamamoto:2007ah}, we distinguish the following phases,
\begin{subequations} \label{phases}
\bea
&{\rm NG \;\, phase:}& \;\; \sigma\neq 0 \, , \;\; d = 0 \, , \\
&{\rm COE \;\, phase:}& \;\; \sigma\neq 0 \, , \;\; d \neq 0 \, .
\eea
\end{subequations} 
In the Nambu-Goldstone (NG) phase there is a chiral condensate but no diquark condensate, while the coexistence (COE)
phase is a color superconductor in the CFL phase with a nonvanishing chiral condensate. One might distinguish
two more cases. First, both condensates may vanish. This was termed the normal (NOR) phase in Ref.\ \cite{Hatsuda:2006ps,Yamamoto:2007ah}.
Clearly, in the presence of the linear $\sigma$ term, we cannot have two vanishing condensates, as we see from Eq.\ (\ref{stat1a}).
Only for $a_0=0$ this case occurs. It thus appears as a special case of the NG phase and is realized on the $a_0=0$ plane in our
three-dimensional $(a_0,a,\alpha)$ phase space. In all other cuts for fixed $a_0$ there can only be an approximate NOR phase with a small 
but nonzero value of $\sigma$. A similar discussion
applies to the phase with nonzero diquark condensate but vanishing chiral condensate, i.e., a ``pure'' CFL phase.
 With Eq.\ (\ref{stat1a}) and $\sigma=0$, the diquark 
condensate becomes
\be
d^2 = \frac{a_0}{\gamma} \, .
\ee
Since we require $d^2>0$, a CFL phase is possible for $a_0>0$. Then, Eq.\ (\ref{stat1b}) yields a constraint for the parameters,
\be
\alpha = -a_0\frac{\beta}{\gamma} \, .
\ee
For a given $a_0$, this is simply a straight line in the $(a,\alpha)$ plane where $\sigma$ vanishes. This line is of no 
particular interest and we obtain it as a limit case of the COE phase. Therefore we can restrict ourselves to the two phases given in 
Eq.\ (\ref{phases}). First we shall discuss them separately in order to determine first-order phase transitions and critical points
within the respective phase. Then we compare the free energies of the phases to obtain the phase diagram.

\subsection{NG phase}

\begin{figure} [t]
\begin{center}
\includegraphics[width=0.5\textwidth]{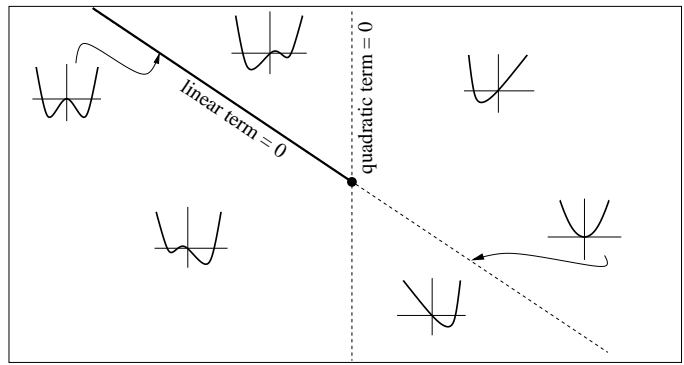}
\caption{Illustration of the appearance of a critical point for a free energy with a linear, quadratic, and positive quartic term in the order 
parameter, such as in Eqs.\ (\ref{OmNG}) and (\ref{OmCOE}). The coefficients
in front of the linear and the quadratic term change their sign across the two straight lines. For a negative quadratic term there is a 
first-order phase transition, indicated by the thick solid line. This line ends at the critical point at which both linear and quadratic terms 
vanish. Across the 
dashed lines there is no phase transition. This implies that one can smoothly connect two 
nontrivial minima of the potential by ``going around'' the first order line. In our context, these two phases are  
the NG phase and the (approximate) NOR phase or the COE phase and the (approximate) CFL phase. For these two cases, 
the location of the critical line and the critical point in the Ginzburg-Landau phase diagram are shown in the left and right panels of 
Fig.\ \ref{figCPs}. 
}
\label{figCP}
\end{center}
\end{figure}

In the NG phase, Eq.\ (\ref{stat1b}) is automatically fulfilled, and Eq.\ (\ref{stat1a}) yields a cubic equation for $\sigma$,
\be \label{cubic1}
0 = a_0+a\sigma - c\sigma^2 + b\sigma^3 \, .
\ee
The general solution of this equation is lengthy and not very interesting. Throughout the paper, the determination of the phase structure
requires solving cubic equations and insertion of these solutions into the free energy. This is best done on a computer 
and hence we do not attempt to derive all details of the phase structure analytically. Many features of the phase diagram, however, can 
be obtained in a simple way and understood on the basis of symmetry arguments as we explain in the following. 

\begin{figure} [t]
\begin{center}
\hbox{\includegraphics[width=0.45\textwidth]{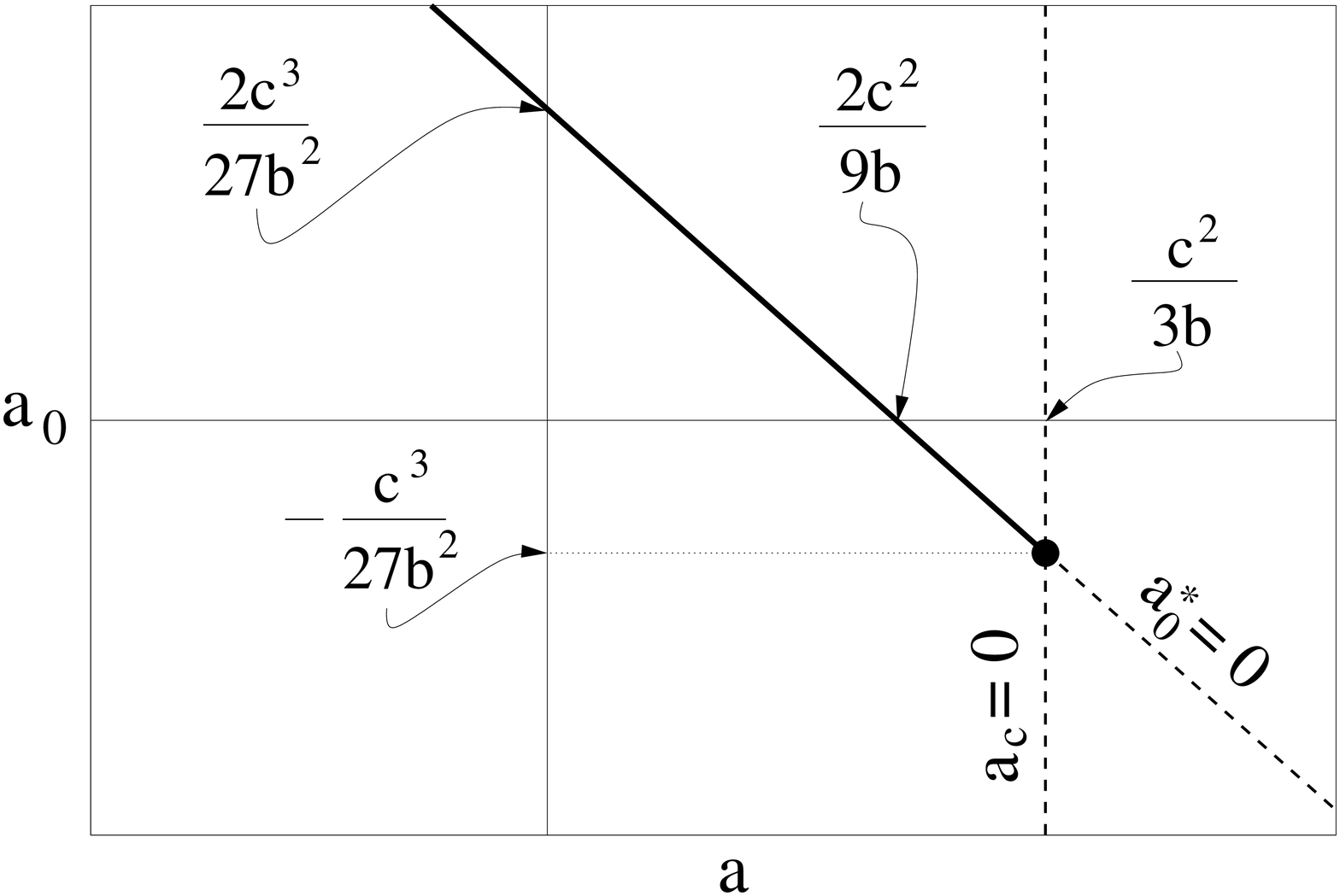}\hspace{0.3cm}
\includegraphics[width=0.45\textwidth]{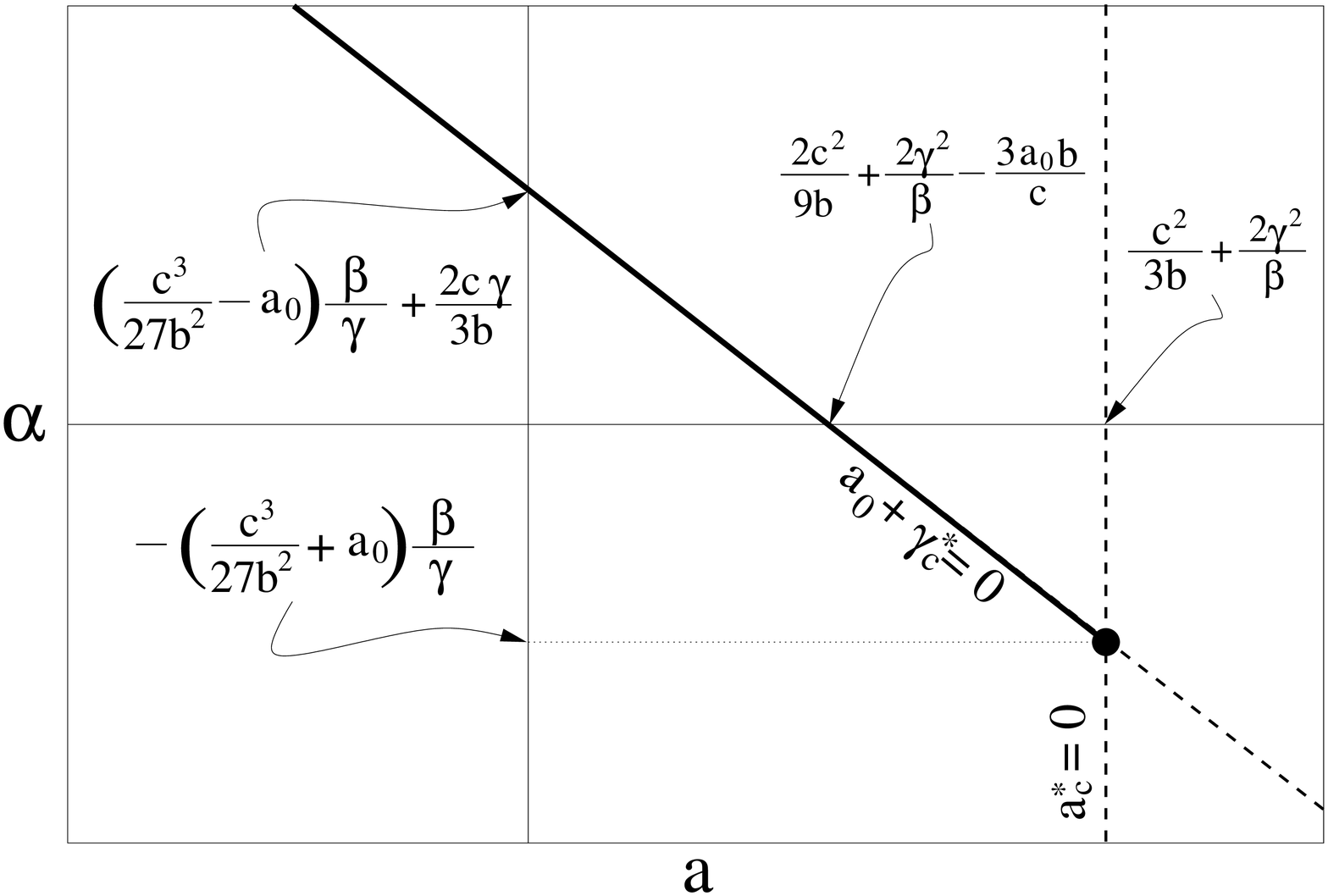}}
\caption{Left panel: first order phase transition line (thick solid line) and critical point in the $(a,a_0)$ plane within the NG phase, 
see Fig.\ \ref{figCP} for a general explanation. The plot shows the coordinates of the critical point
and of the intersections of the lines $a_0^*=0$, $a_c=0$ with the coordinate axes, derived from Eq.\ (\ref{OmNG}).
Since the NG phase is given by the chiral potential $\Omega_{\Phi}$ only, the coordinates depend on the Ginzburg-Landau coefficients
$b$ and $c$, and not on $\alpha$, $\beta$, and $\gamma$. In the 
three-dimensional $(a_0,a,\alpha)$ diagram, the critical point thus becomes a critical line parallel to the $\alpha$-axis where a first-order
phase transition surface ends. Right panel: analogous scenario for the COE phase in the $(a,\alpha)$ plane (note different 
vertical axes of the two plots!), derived from the potential (\ref{OmCOE}). This critical point in the COE phase is present only due to 
instanton effects and vanishes from the phase diagram for $\gamma=0$, leaving behind a first-order line which does not end. }
\label{figCPs}
\end{center}
\end{figure}

By defining 
\be \label{tau}
\tau \equiv \sigma - \frac{c}{3b} \, ,
\ee
we can write the potential of the NG phase as
\bea \label{OmNG}
\Omega_{\rm NG}(\tau) &=& \Omega_0 + a_0^*\tau + \frac{a_c}{2}\tau^2 + \frac{b}{4}\tau^4 \, , 
\eea
such that the cubic term has been eliminated. We have abbreviated
\begin{subequations}
\bea
\Omega_0 &\equiv& \frac{a_0 c}{3b} + \frac{ac^2}{18 b^2} - \frac{c^4}{108 b^3} \, , \\
a_0^* &\equiv& a_0 + \frac{ac}{3b} - \frac{2c^3}{27 b^2} \, , \\
a_c&\equiv& a - \frac{c^2}{3b} \, .
\eea
\end{subequations}
The cubic equation (\ref{cubic1}) then simplifies to 
\be
0 = a_0^* + a_c\tau + b\tau^3 \, .
\ee
The phase structure within the NG phase is now determined by the signs of the linear and quadratic terms in the potential. Therefore,
we need to determine the lines $a_0^*=0$, $a_c=0$. Their intersection represents a critical point at which a first
order phase transition line ends. This is explained for a general potential of the form (\ref{OmNG}) in Fig.\ \ref{figCP}. 
In the left panel of Fig.\ \ref{figCPs} we specify the location of the critical line and the critical point of the NG phase in the $(a,a_0)$ plane 
(the NG phase is obviously independent of $\alpha$).

In the case of the NG phase, the stability criterion from the Hessian (\ref{hessian}) is very simple. The positivity of
its eigenvalues is equivalent to the conditions
\be \label{NGstable}
a> 2c\sigma - 3b\sigma^2 \,, \qquad \alpha > 2\gamma\sigma \, .
\ee
The curve $\alpha=2\gamma\sigma(a)$ in the $(a,\alpha)$ plane with $\sigma(a)$ being the solution to the cubic equation (\ref{cubic1}) 
will turn out to coincide with the second-order phase transition lines between the NG and COE phases, as we demonstrate in Fig.\ \ref{figmass}. 
(The first condition, apart from being a crosscheck for the stability of the NG phase, is of no further interest to us.)

\subsection{COE phase}

For the COE phase we use Eq.\ (\ref{stat1b}) to write the diquark condensate as a function of the chiral condensate,
\be
d^2(\sigma) = \frac{2\gamma\sigma-\alpha}{\beta} \, .
\ee
(In the numerical evaluation, one has to use this relation to check whether for a given solution $\sigma$, the diquark condensate is
real, $d^2>0$.) Inserting this into the potential (\ref{Om0}) yields
\be \label{OmCOE1}
\Omega_{\rm COE}[\sigma,d^2(\sigma)] = -\frac{\alpha^2}{4\beta} + (a_0+\gamma^*)\sigma +\frac{a^*}{2}\sigma^2 -\frac{c}{3}\sigma^3
+\frac{b}{4}\sigma^4 \, ,
\ee
where we have used the notation from Ref.\ \cite{Yamamoto:2007ah},
\be\label{gammastar}
\gamma^* \equiv \frac{\alpha\gamma}{\beta} \, ,\qquad a^*\equiv a - \frac{2\gamma^2}{\beta} \, .
\ee
We can now proceed as for the NG phase. With the new variable $\tau$ from Eq.\ (\ref{tau}) we can write the potential as
\bea \label{OmCOE}
\Omega_{\rm COE}(\tau) &=& \frac{a_0 c}{3b} + \Omega_c  + (a_0+\gamma_c^*) \tau + \frac{a_c^*}{2}\tau^2 + \frac{b}{4}\tau^4 \, , 
\eea
where all $a_0$'s are written explicitly. A nonzero $a_0$ gives additional contributions to the constant and linear terms in $\tau$.
All other terms are identical to the massless case \cite{Yamamoto:2007ah} (remember that we have absorbed the mass
terms in the overall coefficients unless they have produced new structures in the order parameters),
\begin{subequations}
\bea
\Omega_c &\equiv & -\frac{\alpha^2}{4\beta} + \frac{\gamma^*c}{3b} + \frac{a^*c^2}{18b^2} - \frac{c^4}{108b^3} \, , \\
\gamma_c^*&\equiv& \gamma^* + \frac{a^* c}{3b}-\frac{2c^3}{27b^2} \, , \\
a_c^*&\equiv& a^* - \frac{c^2}{3b} \, .
\eea
\end{subequations}
The stationarity equation is thus 
\be \label{cubic3}
0 = a_0+\gamma_c^* + a_c^*\tau + b\tau^3 \, .
\ee
Again we can determine the transition line for the first-order transition within the COE phase and the corresponding critical point.
We show the coordinates of this line in the right panel of Fig.\ \ref{figCPs}. 
It is obvious that the critical point is a consequence of the anomaly: if the anomalous term vanishes, $\gamma=0$, the critical 
point disappears from the phase diagram because its $\alpha$ coordinate goes to $-\infty$ or $+\infty$, depending on the  
sign of $\frac{c^3}{27b^2}+a_0$ (while its $a$ coordinate remains finite). 
We also see that a negative $a_0$ shifts, for a given anomaly coefficient $\gamma$, the critical point towards larger values
of $\alpha$ while keeping the $a$ coordinate fixed. Remember that for both plots in Fig.\ \ref{figCPs} we only know that the respective phases 
are a stationary point of the potential, and we have not yet determined the global minimum. And indeed, for negative $a_0$ with 
sufficiently large modulus, the $\alpha$ coordinate of the critical point is shifted into region where the NG phase, not the COE phase, 
is the ground state, i.e., the critical point has disappeared. We demonstrate this scenario in the next subsection in Fig.\ \ref{figmass}.

\begin{figure} [t]
\begin{center}
\hbox{\includegraphics[width=0.5\textwidth]{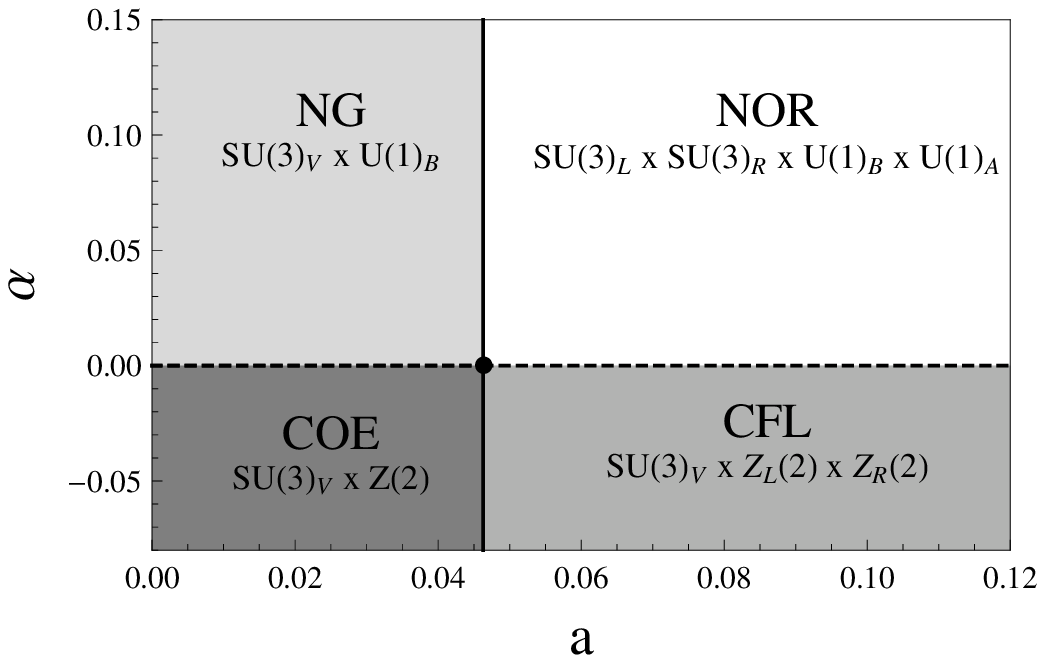}
\includegraphics[width=0.5\textwidth]{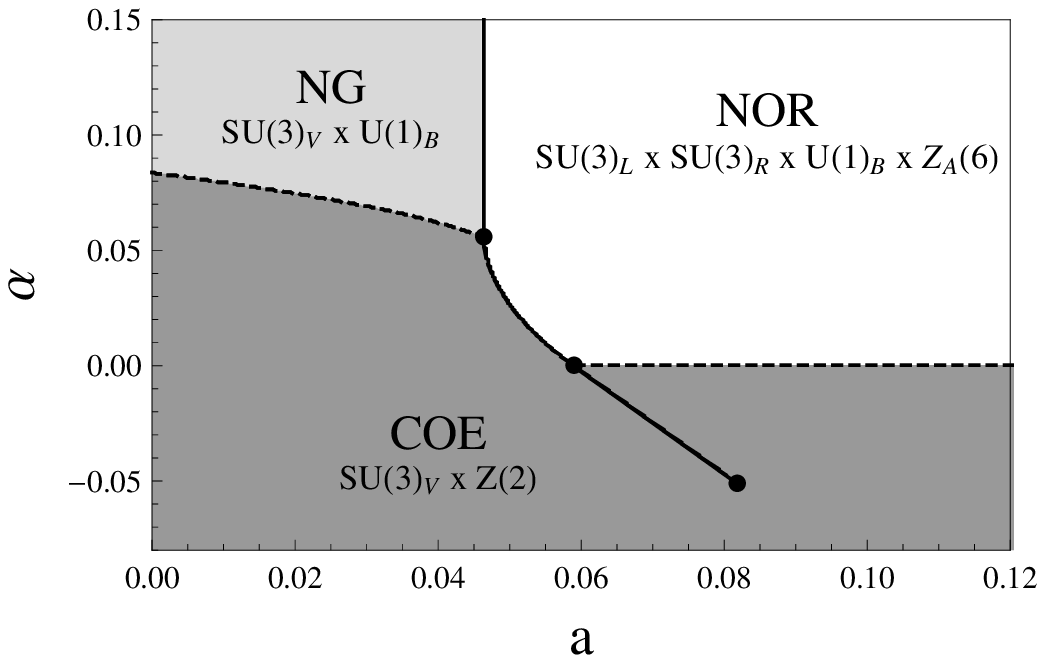}}
\caption{Phase diagrams in the $(a,\alpha)$ plane without strange quark mass effects and without meson condensation, without (left panel)
and with (right panel) anomalous effects. First-order phase transition lines are solid, 
second-order lines are dashed. For each phase we have indicated its global symmetries. They 
explain why the first-order phase transition line does not (left) and does (right) end at a critical point (see text and Fig.\ \ref{figdiscrete}
for detailed explanations).
The vertical (first-order) phase transition line separating NG from NOR (left and right) and COE from CFL (left) is located at $a=2c^2/(9b)$. 
For the coordinates of the critical point (right) see right panel of Fig.\ \ref{figCPs}. 
We have set the Ginzburg-Landau parameters to $b=1.2$, $c=0.5$, $\beta=1.6$ for both plots, and $\gamma=0$ (left), $\gamma=0.1$ (right). 
This particular choice is completely irrelevant for the topology of the left plot. For the right plot, one slightly different topology can 
be obtained in a different class of parameter values, where the phase transition between NG and COE is of first order along a certain piece of the 
transition line, see right panel of Fig.\ 2 in Ref.\ \cite{Yamamoto:2007ah}. We translate the scenarios shown here into possible 
topologies of the QCD phase diagram in Fig.\ \ref{figQCD}. 
}
\label{fignomass}
\end{center}
\end{figure}

\subsection{Ginzburg-Landau phase diagram, symmetries, and conjectured QCD phase diagram}

The next step is to compare the free energies of the NG and COE phases. We start with the case $a_0=0$
in order to reproduce the results from Refs.\ \cite{Hatsuda:2006ps,Yamamoto:2007ah}. We refer the reader to appendix A of 
Ref.\ \cite{Yamamoto:2007ah} for an analytical derivation of some of the phase transition lines. 
The results without and with anomalous term are shown in Fig.\ \ref{fignomass}.

\begin{figure} [t]
\begin{center}
\includegraphics[width=\textwidth]{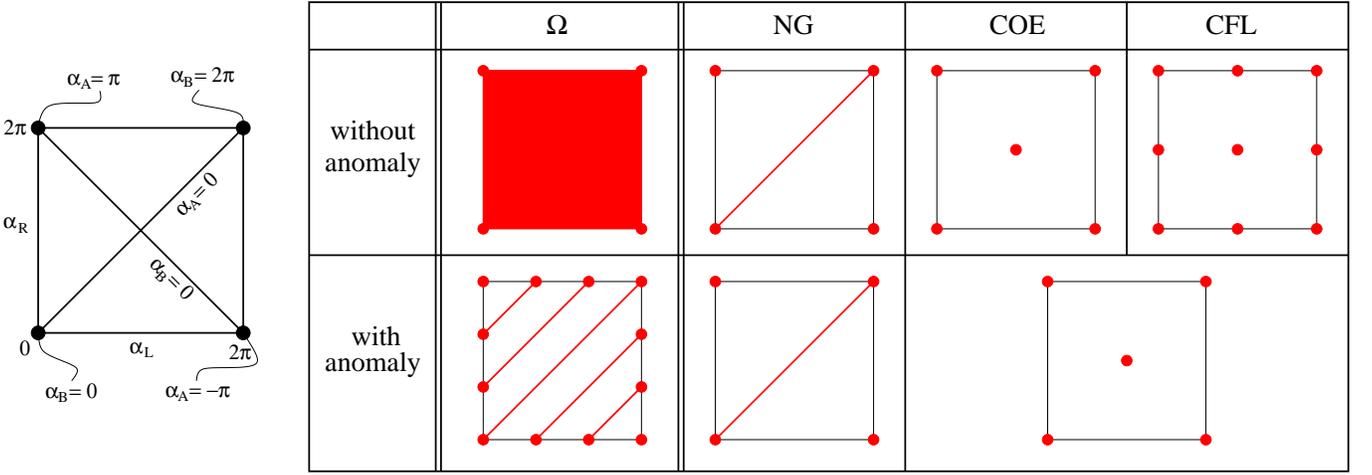}
\caption{(Color online) Diagram (left): topological space of $U(1)_L\times U(1)_R$ with generators $\alpha_L,\alpha_R\in [0,2\pi]$. Opposite 
sides of the square have to be identified (in particular, all four corners of the square correspond to the unit element). 
The resulting torus can also be parametrized by 
$\alpha_B=(\alpha_R+\alpha_L)/2 \in [0,2\pi]$ and $\alpha_A=(\alpha_R-\alpha_L)/2 \in [-\pi,\pi]$. Table (right): symmetries of the 
free energy $\Omega$ and the NG, COE, and CFL phases without and with axial anomaly. Thick (red) lines and points indicate group 
elements under which the free energy and the respective phases are invariant. For instance, in the first row, the points for the COE phase are
obtained as the common subset of the points of the NG and CFL phases (and all sets of points are subsets of the points for $\Omega$). 
Only with anomaly, i.e., only after restricting the symmetry of $\Omega$ from $U(1)_L\times U(1)_R$ to $U(1)_B\times \mathbb{Z}_A(6)$, 
the symmetries of COE and CFL are identical, allowing for a smooth crossover between these two phases. 
}
\label{figdiscrete}
\end{center}
\end{figure}

Since we have set the linear term in the chiral potential to zero, there is a NOR phase in both scenarios of the figure, 
where the chiral condensate vanishes exactly.
If we switch off the anomaly, $\gamma=0$, the interaction term $\propto d^2\sigma$ vanishes. Since this is the only interaction term in our 
approximation, the chiral and diquark parts of the free energy decouple, resulting in a very simple phase diagram (left panel). 
In this phase diagram COE and CFL phases are separated by a first order line which does not end at a critical point (vertical solid line). 
This is a consequence of the global symmetries of the two phases. As indicated in the figure, the symmetries of COE and CFL are 
different without the anomaly. In other words, adding a chiral condensate 
to the CFL phase changes the (discrete) symmetries of the phase. This can be seen from the transformation properties of the order parameters in 
Eqs.\ (\ref{Phitrafo}) and (\ref{dLdRtrafo}). The chiral condensate $\Phi={\rm diag}(\sigma,\sigma,\sigma)$ and the CFL diquark 
condensate spontaneously break the chiral group $SU(3)_L\times SU(3)_R$ down to the 
vector subgroup of simultaneous left- and right-handed rotations $SU(3)_V$. From Eq.\ (\ref{Phitrafo}) we see that any chiral condensate 
is also invariant under $U(1)_B$ transformations and under transformations of an axial subgroup $\mathbb{Z}_A(2)$.
However, this $\mathbb{Z}_A(2)$ is contained in $U(1)_B$. To see this it is helpful to consider the group
$U(1)_L\times U(1)_R$ as a topological space, in this case a torus, on which the discrete subgroups are sets of discrete points. We show
this geometric picture in Fig.\ \ref{figdiscrete} which illustrates and facilitates the discussion of the discrete subgroups, in particular
since we switch repeatedly between the bases of left- and right-handed rotations vs.\ axial and vector rotations, which can be confusing 
without this illustration.

In contrast to the chiral condensate, the diquark condensates $d_L$, $d_R$ break $U(1)_B$ spontaneously. Hence there must be a true phase 
transition separating the NG and NOR phases from the COE and CFL phases. This statement is independent of the anomaly. The anomaly becomes important
for the difference between COE and CFL phases. In the absence of instanton effects, the diquark condensates are invariant
under independent sign flips of left- and right-handed quark fields, $\mathbb{Z}_L(2)\times \mathbb{Z}_R(2)$. This discrete group is broken by the 
chiral condensate; therefore, the COE phase, containing both order parameters, is only invariant under the common subgroup of $U(1)_B$ and 
$\mathbb{Z}_L(2)\times \mathbb{Z}_R(2)$.
This is the group of simultaneous sign flips, $\mathbb{Z}(2)$. As a result, without axial anomaly COE and CFL have different residual 
symmetry groups, see first row of the table in Fig.\ \ref{figdiscrete}.

The anomaly reduces the axial symmetry $U(1)_A$ of the potential to $\mathbb{Z}_A(6)$, see Eq.\ (\ref{anomtrafo}). 
The group $U(1)_B\times\mathbb{Z}_A(6)$ is 
represented in the first panel of the second row in Fig.\ \ref{figdiscrete}. The anomaly does not affect the residual group of the NG phase. 
However, the residual group of the CFL phase is reduced since only $\mathbb{Z}(2)\subset \mathbb{Z}_L(2)\times \mathbb{Z}_R(2)$, not the
entire group $\mathbb{Z}_L(2)\times \mathbb{Z}_R(2)$, is a subgroup
of $U(1)_B\times\mathbb{Z}_A(6)$, as can be seen geometrically in Fig.\ \ref{figdiscrete}. 
Therefore, if instanton effects are taken into account, CFL is invariant only under the group
of simultaneous sign flips $\mathbb{Z}(2)$. Now the addition of the chiral condensate does not further reduce this group, and the 
residual groups of COE and CFL become identical, see second row of Fig. \ref{figdiscrete}. This allows for a smooth crossover between these two
phases, and thus the first-order line between COE and (approximate) CFL can end at a critical point. 

This expectation from symmetry arguments is borne out in the Ginzburg-Landau phase diagram, see right panel of Fig.\ \ref{fignomass}. 
This diagram will serve as a basis for our extensions in the following. 

One might ask why the straight line of the first order phase transition in the COE 
phase terminates at the intersection with the second-order
line and then continues as a nontrivial, non-straight, curve. From the right panel of Fig.\ \ref{figCPs} one could have expected the 
straight first-order line
to be present wherever the $a_0+\gamma_c^*=0$ line lives in the COE region. The reason for this behavior is the condition $d^2>0$. 
In terms of the schematic potentials of Fig.\ \ref{figCP}, the first order line separates two nontrivial minima. Here, for positive $\alpha$, 
the second minimum is forbidden since it would imply $d^2<0$. Consequently, upon crossing the line $a_0+\gamma_c^*=0$ the ground state remains 
in the same local minimum although, if one had ignored the condition $d^2>0$, there would have be a second local minimum with lower free energy.

\begin{figure} [t]
\begin{center}
\hbox{\includegraphics[width=0.5\textwidth]{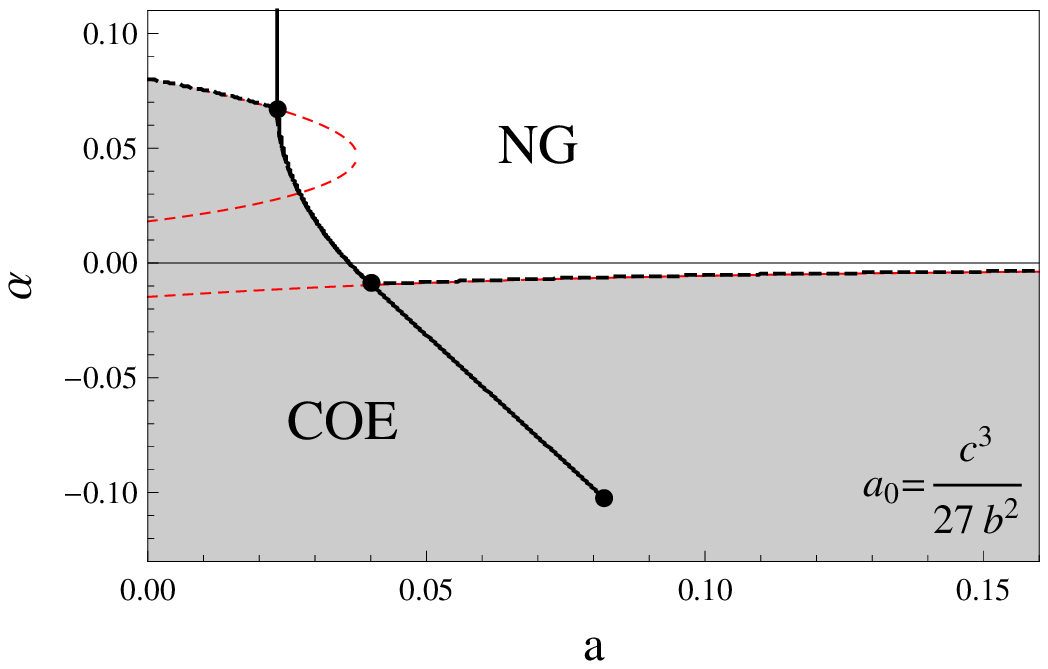}
\includegraphics[width=0.5\textwidth]{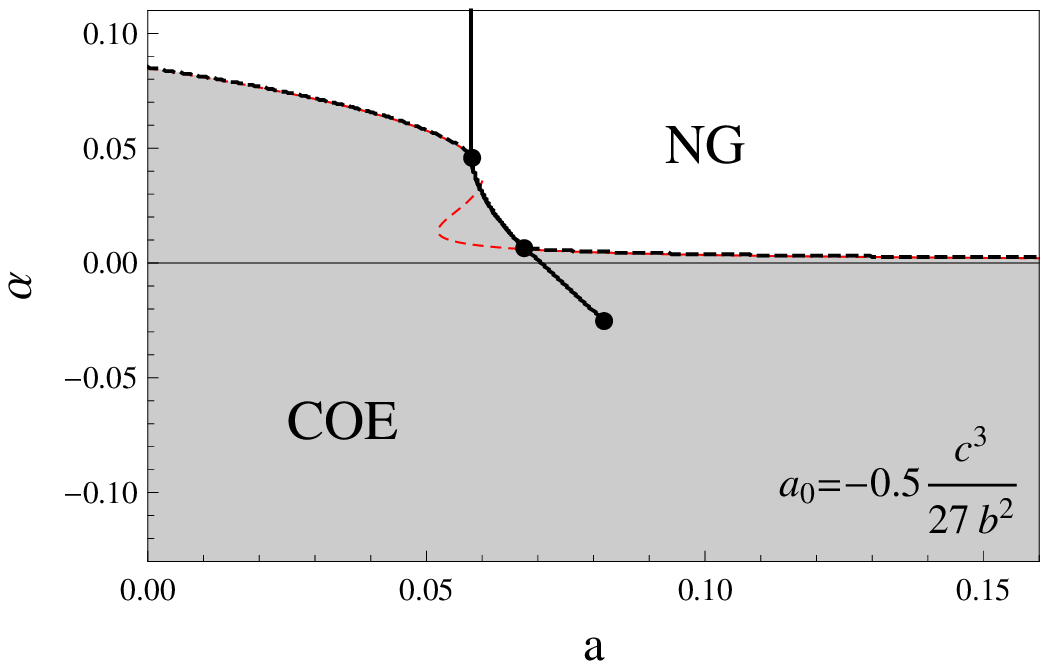}}

\hbox{\includegraphics[width=0.5\textwidth]{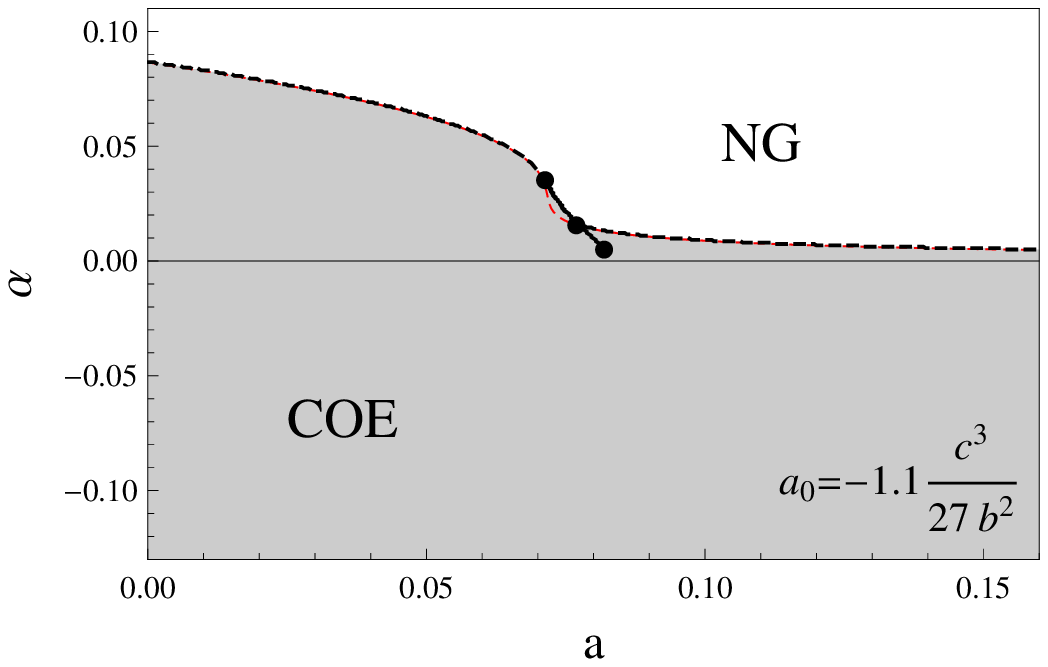}
\includegraphics[width=0.5\textwidth]{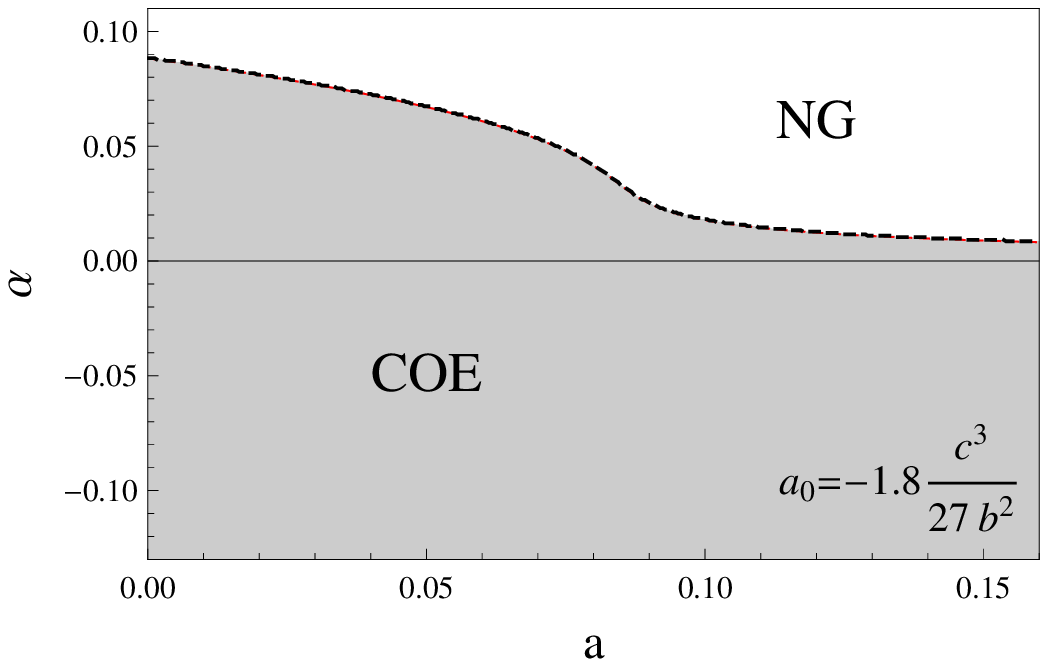}}
\caption{(Color online) Ginzburg-Landau phase diagrams in the $(a,\alpha)$ plane with quark mass effect in the chiral potential 
(without meson condensation). Solid lines are first-order phase transitions, thick (black) dashed lines are second-order phase transitions. 
We have chosen various values of the linear coefficient $a_0$ of the chiral condensate and fixed all other parameters as in the right panel 
of Fig.\ \ref{fignomass} (in particular, $\gamma=0.1$). The parameter $a_0$ is given in units of
$\frac{c^3}{27b^2}$ because for $a_0<-\frac{c^3}{27b^2}$ the critical line which separates NG and approximate NOR phases vanishes and
there is a crossover between these two phases. 
The thin (red) dashed line is the curve $\alpha=2\gamma \sigma(a)$ with $\sigma(a)$ given by the solutions of the cubic equation (\ref{cubic1}).
This curve is obtained from the stability criterion of the NG phase and coincides with all second-order phase transition lines. 
}
\label{figmass}
\end{center}
\end{figure}

Let us now switch on the effect of the linear term in $\sigma$, induced by a nonzero strange quark mass. The chiral group 
$SU(3)_L\times SU(3)_R$ is now only approximate, which allows for a smooth crossover between the NG and (approximate) NOR phases, see
left panel of Fig.\ \ref{figCPs}. More precisely, within our ansatz $M={\rm diag}(0,0,m_s)$ we leave part of the chiral group intact,
namely the $SU(2)_L\times SU(2)_R$ associated with the $u$ and $d$ quarks. On the other hand, our ansatz  
$\Phi={\rm diag}(\sigma,\sigma,\sigma)$ includes chiral condensates for the $u$ and $d$ quarks and thus one might think that $\Phi$ {\it does} 
break the chiral symmetry in the $u$, $d$ sector 
spontaneously. However, as one can see for instance from Eq.\ (\ref{phiM}), once we have set $\sigma_u=\sigma_d=\sigma_s$  
it makes no difference (up to a numerical prefactor which is unimportant for our purpose) whether we set $m_u=m_d=0$ or not. 
Therefore, although we consider the simplified situation of vanishing $u$ and $d$ quark masses, the NG and NOR phases are no longer 
distinguished by symmetry in the presence of a nonzero $m_s$. 
As a consequence, we expect the first-order line between NG and NOR in the $a_0=0$ plane to become a first-order surface in the 
three-dimensional ($a_0,a,\alpha)$ phase diagram which ends at a critical line at $a_0=-\frac{c^3}{27b^2}$. In other words, if we plot
phase diagrams in the $(a,\alpha)$ plane for fixed values of $a_0$, this first order line should be absent for all $a_0<-\frac{c^3}{27b^2}$. 
The numerical evaluation confirms this expectation, see Fig.\ \ref{figmass}. Besides the phase transition lines we 
have also plotted the curve $\alpha=2\gamma \sigma(a)$ (red dashed line), where $\sigma(a)$ is given by the real solution of Eq.\ (\ref{cubic1}).
This curve is obtained from the stability criterion of the NG phase, see Eq.\ (\ref{NGstable}). 
The main purpose of this curve is to provide a semi-analytical form of the second-order phase transition lines. In Fig.\ \ref{figmass}
they have been obtained independently by directly comparing the free energies of the different phases.

Besides the disappearance of the critical line between NG and approximate NOR phases (and deformations of the transition lines 
which do not change the topology of the phase diagram and thus are not very interesting for our purpose)
there is one more topological change upon varying $a_0$. Namely, we see that for sufficiently small values of $a_0$ the first-order 
line within the COE phase disappears and the phase diagram in the $(a,\alpha)$ plane consists of a sole second-order transition separating 
NG and COE phases. More precisely, upon decreasing $a_0$ the critical point moves towards larger values of $\alpha$ while its
$a$ coordinate remains fixed. Interestingly, at the same value of $a_0$ where the first order line between NG and approximate NOR phases disappears,
the critical point sits on the $\alpha=0$ axis. We can see this analytically from Fig.\ \ref{figCPs}. Then, upon further decreasing $a_0$, 
it approaches the phase transition line between NG and COE phases and disappears for values of $a_0$ below some critical value for which we do 
not have an analytic expression. We have thus found two different ways to make the critical point disappear: switching off the anomaly removes
the critical point but leaves the first-order critical line, while going to (possibly unphysically) small values of $a_0$ removes the 
critical point {\em and} the critical line. 

\begin{figure} [t]
\begin{center}
\includegraphics[width=\textwidth]{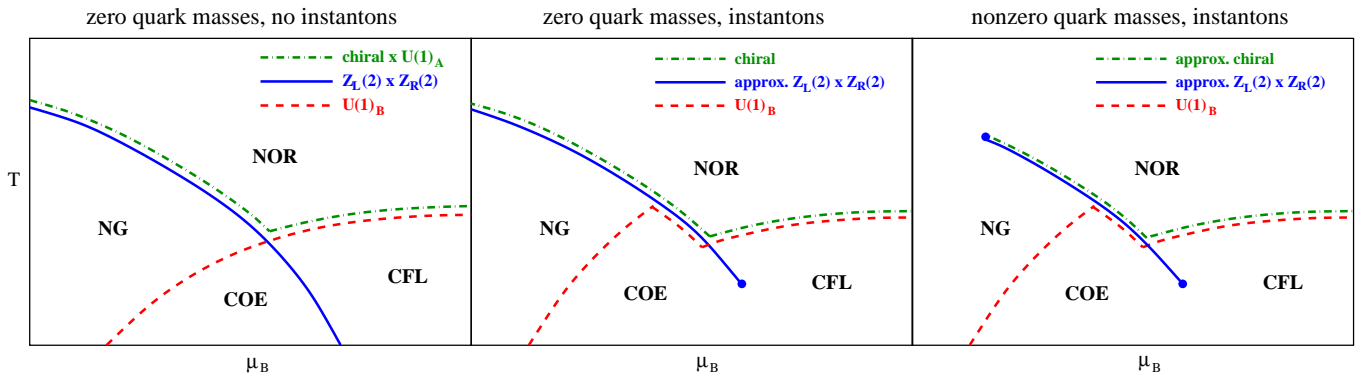}
\caption{(Color online) Conjectured translations of the Ginzburg-Landau phase diagrams to the QCD phase diagram in the $(\mu_B,T)$ 
plane \cite{talk}. We have indicated the global symmetries which are broken in the 
various transitions. Left panel: zero quark masses, no instanton effects,   
corresponding to the left panel in Fig.\ \ref{fignomass}. Middle panel: zero quark masses, nonzero instanton effects. In this case
the CFL and COE phases are no longer distinguished by symmetry and thus allow for a smooth crossover. Whether the critical
point is indeed present in the QCD phase diagram cannot be decided from the Ginzburg-Landau study; the $T=0$ axis may or 
may not intersect with the first order line in the COE phase in the right panel of Fig.\ \ref{fignomass}. Here we show the case where 
it does not. Right panel: nonzero quark masses, nonzero instanton effects, see Fig.\ \ref{figmass}. Here we know from lattice QCD 
that the first-order phase transition between NG and (approximate) NOR does not reach the $\mu_B=0$ axis.  
}
\label{figQCD}
\end{center}
\end{figure}

It is interesting to speculate how the Ginzburg-Landau results translate into the QCD phase diagram. For a precise translation our simple approach 
is of course not sufficient. Firstly, the Ginzburg-Landau potential is an expansion in the order parameters and thus cannot account for the 
complete potential; the expansion is expected to fail far away from second-order phase transitions where fluctuations of the phase of the 
order parameter can become more important than fluctuations of the magnitude of the order parameter. We have also neglected gauge field fluctuations
which in fact drive the transition from a color-superconducting to the normal phase first order \cite{Matsuura:2003md,Giannakis:2004xt}. 
Secondly, even if we assume the Ginzburg-Landau approximation to be valid 
for all temperatures and densities of interest, we would need the dependence of the Ginzburg-Landau coefficients on the baryon
chemical potential $\mu_B$ and temperature $T$. This 
dependence is not known within full QCD, since the relevant regions of the phase diagram involve strong-coupling effects, and lattice
calculations are inapplicable due to the sign problem at finite $\mu_B$. We may still conjecture a translation from the Ginzburg-Landau phase 
diagrams to QCD, see Fig.\ \ref{figQCD}. In this figure, the left and middle panel correspond to the situation without strange quark mass,
and thus to the Ginzburg-Landau diagrams in Fig.\ \ref{fignomass}.

For the interpretation of our Ginzburg-Landau results with strange quark mass, it is helpful to think
of the QCD $(\mu_B,T)$ plane to be a complicated surface in our $(a_0,a,\alpha)$ parameter space. We know from lattice calculations
that, at $\mu_B=0$, the transition from the chirally broken to the chirally (approximately) symmetric phase is a smooth crossover
\cite{Aoki:2006we}. Therefore, 
the temperature axis must not intersect the critical surface between NG and (approximate) NOR phases. 
This critical surface may then manifest itself as a critical line between NG and approximate NOR phases which 
ends at a critical point, see right panel of Fig.\ \ref{figQCD}. Another logical possibility is the absence of this line \cite{deForcrand:2008vr}, 
which would for instance be realized if the whole $(\mu_B,T)$ surface were located ``behind'' the $a_0=-\frac{c^3}{27b^2}$ plane.

For the critical point at low $T$ and large $\mu_B$ our analysis with 
finite strange quark mass has opened up a third possibly topology. Without mass, the point may, although always present
in the Ginzburg-Landau phase diagram, be either outside the $(\mu_B,T)$ plane (formally, one can think of the point being located at negative $T$)
or within the $(\mu_B,T)$ plane. With mass effect we have seen that the critical line that ends at this critical point may be absent in the 
Ginzburg-Landau phase diagram. Thus, if the $(\mu_B,T)$ surface is located at sufficiently negative $a_0$, there is no 
first-order transition within the COE phase.

\section{Including meson condensation}
\label{sec:K0}

We can now extend the results from the previous section by allowing for a nonzero kaon condensate $\phi$. In principle, this requires
to consider several additional independent Ginzburg-Landau parameters as we can see from the full potential 
(\ref{Omfull}). In this potential, the parameters $\alpha_2$, $\beta_2$, $\mu^2$, and $\gamma_2$ become relevant when we allow for nontrivial 
values of $\phi$. All of these parameters correspond to mass terms and one might, as a first approximation, neglect these terms. However, then 
the only nontrivial structure involving $\phi$ is the unsuppressed $d^2\sigma\cos\phi$ term, and there would be no kaon
condensation at all (in principle, there could be a condensate at the fixed value of $\phi=\pi$). Therefore, we have to 
keep the $\mu^2\sin^2\phi$ term in order to match our potential to the high-density effective theory, see discussion in Sec.\ \ref{sec:diquark}, 
but for simplicity neglect the terms proportional to $\alpha_2$, $\beta_2$, and $\gamma_2$. We have checked numerically that the inclusion of 
these terms can indeed make a difference to the topology of the phase diagrams with kaon condensate. We shall come back to this issue 
in the discussion at the end of Sec.\ \ref{sec:2SC}.

Within this approximation, the only additional parameter compared to the previous section is $\mu^2$, and our potential becomes
\bea \label{OmK0}
\Omega(\sigma,d,\phi) &=& a_0\sigma + \frac{a}{2}\sigma^2-\frac{c}{3}\sigma^3 +\frac{b}{4}\sigma^4 + \frac{\alpha}{2} d^2 
+ \frac{\beta-\mu^2\sin^2\phi}{4} d^4  -\gamma\frac{1+2\cos\phi}{3} d^2\sigma \, .
\eea
A comparison with the potential of the high-energy effective theory (\ref{Vphi}) shows that one can consider the interaction term 
$d^2\sigma\cos\phi$ as an effective, dynamical mass term for the kaon. 
The boundedness of the potential requires the $d^4$ term to be positive, which yields an upper bound for $\mu^2$,
\be
\mu^2< \beta \, .
\ee
The stationarity equations (\ref{stat}) are  
\begin{subequations} \label{stat2}
\bea
0 &=& \frac{\partial\Omega}{\partial \sigma} = a_0 + a\sigma - c\sigma^2 + b\sigma^3 - \gamma \frac{1+2\cos\phi}{3} d^2 \, , \label{stat2a}\\
0 &=& \frac{\partial\Omega}{\partial d} = \alpha d + (\beta-\mu^2\sin^2\phi) d^3 - 2\gamma \frac{1+2\cos\phi}{3} d \sigma \, , \label{stat2b}\\
0 &=& \frac{\partial\Omega}{\partial \phi} = d^2\sin\phi\left[\frac{2\gamma}{3}\sigma-\frac{\mu^2}{2}d^2\cos\phi\right] \, .\label{stat2c}
\eea
\end{subequations}
We distinguish the following phases, 
\begin{subequations} \label{phases2}
\bea
&{\rm NG \;\, phase:}& \;\; \sigma\neq 0 \, , \;\; d = 0 \, , \label{NGphase2}\\
&{\rm COE \;\, phase:}& \;\; \sigma\neq 0 \, , \;\; d \neq 0 \, , \;\; \phi=0 \, ,\\
&\mbox{COE-$K^0$  phase:}& \;\; \sigma\neq 0 \, , \;\; d,\phi \neq 0 \, .
\eea
\end{subequations} 
The first two phases are the same as in the previous section. Note that $\phi$ only appears in the potential when $d$ is nonzero. This
is clear since without diquark condensation there are no kaons to condense. Therefore, the NG phase does not depend on $\phi$, and we
have not specified its values in Eq.\ (\ref{NGphase2}). The value of $\phi$ distinguishes between the phases with nonzero CFL order parameter,
COE and COE-$K^0$. Again, as discussed for the case without meson condensate below Eq.\ (\ref{phases}), the NOR and ``pure'' CFL/CFL-$K^0$ 
phases are obtained as special cases from the NG and COE/COE-$K^0$ phases. They only exist in a two-dimensional subspace of the 
three-dimensional $(a_0,a,\alpha)$ parameter space and thus shall not be further discussed in the following.

\subsection{COE-$K^0$ phase}

In order to compute the phase structure in the presence of a kaon condensate, we need to compute the free energies of the three phases 
(\ref{phases2}). The results for the NG and COE phases can be taken from the previous section. Thus we only have to discuss the 
COE-$K^0$ phase. It is
convenient to express $\phi$ and $d$ as functions of $\sigma$. Solving Eq.\ (\ref{stat2c}) for $\cos\phi$ and inserting the result into 
Eq.\ (\ref{stat2b}) yields 
\begin{subequations} \label{cosphid}
\bea
\cos\phi(\sigma) &=& \frac{4\gamma\sigma}{3\mu^2d^2(\sigma)} \, , \\ 
d^2(\sigma) &=& \frac{2\gamma\sigma-3\alpha}{3(\beta-\mu^2)} \, . \label{dsqsigma}
\eea
\end{subequations}
These expressions can be inserted into Eq.\ (\ref{stat2a}) to obtain an equation for $\sigma$. Equivalently, we can insert them into the 
potential (\ref{OmK0}) and then minimize it with respect to $\sigma$.  The potential becomes 
\be
\Omega_{{\rm COE}-K^0}[\sigma,d(\sigma),\phi(\sigma)] = -\frac{\alpha^2}{4(\beta-\mu^2)}+(a_0+\gamma^*_\mu)\sigma 
+ \frac{a_\mu^*}{2}\sigma^2-\frac{c}{3}\sigma^3 +\frac{b}{4}\sigma^4 \, ,
\ee
where 
\be
\gamma^*_\mu \equiv \frac{\alpha\gamma}{3(\beta-\mu^2)} \, ,\qquad a^*_\mu\equiv a - \frac{2\gamma^2}{9}\left(\frac{1}{\beta-\mu^2}+\frac{4}{\mu^2}
\right) \, .
\ee
We see that the potential of the COE-$K^0$ phase has the same structure as the one of the COE phase (\ref{OmCOE1}),
with modified coefficients $\gamma^*_\mu$, $a^*_\mu$. We may thus proceed analogously to determine the critical point within the 
COE-$K^0$ phase. With the variable $\tau=\sigma-c/(3b)$ from Eq.\ (\ref{tau}) we obtain the potential
\bea \label{OmCOEK0}
\Omega_{{\rm COE}-K^0}(\tau) &=& \frac{a_0 c}{3b} + \Omega_{c,\mu}  + (a_0+\gamma_{c,\mu}^*) \tau + \frac{a_{c,\mu}^*}{2}\tau^2 
+ \frac{b}{4}\tau^4 \, , 
\eea
with
\begin{subequations}
\bea
\Omega_{c,\mu} &\equiv & -\frac{\alpha^2}{4(\beta-\mu^2)} + \frac{\gamma_\mu^*c}{3b} + \frac{a_\mu^*c^2}{18b^2} - 
\frac{c^4}{108b^3} \, , \\
\gamma_{c,\mu}^*&\equiv& \gamma_\mu^* + \frac{a_\mu^* c}{3b}-\frac{2c^3}{27b^2} \, , \\
a_{c,\mu}^*&\equiv& a_\mu^* - \frac{c^2}{3b} \, ,
\eea
\end{subequations}
in complete analogy to Eq.\ (\ref{OmCOE}) and resulting in the stationarity equation 
\bea \label{tauCOEK0}
0= a_0+\gamma_{c,\mu}^*+ a_{c,\mu}^*\tau + b\tau^3 \, . 
\eea
After solving this equation, one has to insert the real solution(s) back into 
Eqs.\ (\ref{cosphid}) to check whether $d^2(\sigma)>0$ and $-1<\cos\phi(\sigma)<1$. If one or both of these conditions are violated, the
given solution for $\sigma$ has to be discarded. 

Since the cubic equation has the same structure as for the COE phase, we obtain an analogous first-order line as shown in the right panel of 
Fig.\ \ref{figCPs}.
From $a_0+\gamma_{c,\mu}^*=a_{c,\mu}^*=0$ we can compute the location of the critical point. Its coordinates turn out to be
\be \label{K0crit}
a_{{\rm COE}-K^0} = \frac{c^2}{3b}+\frac{2\gamma^2}{9}\left(\frac{1}{\beta-\mu^2}+\frac{4}{\mu^2}
\right) \, , \qquad
\alpha_{{\rm COE}-K^0} = -\frac{3(\beta-\mu^2)}{\gamma}\left(a_0+\frac{c^3}{27b^2}\right) \, . 
\ee
We can compare this to the critical point in the COE phase whose coordinates are read off from the right panel of Fig.\ \ref{figCPs},
\be \label{COEcrit}
a_{{\rm COE}} = \frac{c^2}{3b}+\frac{2\gamma^2}{\beta} \, , \qquad
\alpha_{{\rm COE}} = -\frac{\beta}{\gamma}\left(a_0+\frac{c^3}{27b^2}\right) \, . 
\ee
Before we turn to the phase diagram we can give a semi-analytical
expression for the phase transition line between the COE and COE-$K^0$ phases. Since this phase transition will turn out to be of second order, 
we consider nonzero, but very small values of $\phi$, i.e., we approach the second-order phase transition 
from within the COE-$K^0$ phase. Hence we can divide Eq.\ (\ref{stat2c}) by $\sin\phi$ and then set $\cos\phi=1$ to get 
a simple relation between $\sigma$ and 
$d$. With the help of Eq.\ (\ref{dsqsigma}), we eliminate $d$ from this relation and obtain the value of $\sigma$ at the phase boundary 
between COE and COE-$K^0$,
\be \label{sigmacr}
\sigma = \frac{3\alpha}{2\gamma}\frac{\mu^2}{3\mu^2-2\beta} \, .
\ee
Using the relation between $\sigma$ and $\tau$ from Eq.\ (\ref{tau}) we insert this into the stationarity equation (\ref{tauCOEK0})
which then only contains the Ginzburg-Landau parameters. One obtains a cubic equation for $\alpha$ which has a very lengthy, but
analytical, solution $\alpha(a)$ for the second-order phase transition line between COE and COE-$K^0$ phases.

Our next goal is to compute the phase diagram including meson condensation. It is useful to start with the following two questions.
\begin{enumerate}[$(i)$]
\item What is the fate of the critical point in the $(a,\alpha)$ phase diagram in the presence of kaon condensation? We have seen above that 
a first-order line ending at a critical 
point is possible in the COE and in the COE-$K^0$ phase. More precisely, there is a ``would-be'' critical point with coordinates given in 
Eq.\ (\ref{COEcrit}) which, if the COE phase is the ground state at this point, is a true critical point,  and there is a ``would-be''
critical point with coordinates given in Eq.\ (\ref{K0crit}) which, if the COE-$K^0$ phase is the ground state at this point, is a true
critical point. This leaves us with four logical possibilities. There may be no critical point at all when both ``would-be''
critical points are covered by the ``wrong'' phases; there might be one critical point, either in the COE or COE-$K^0$ phase; or 
both ``would-be'' critical points are realized if they are covered by the ``right'' phases. This classification is very useful  
since the information about the critical points determines, to a large extent, the topology of the entire phase diagram. 

\item Is there a region in the parameter space (here we mean all parameters except for $a$ and $\alpha$) for which the COE phase
is completely replaced by the COE-$K^0$ phase in the $(a,\alpha)$ plane? This question is interesting in view of the quark-hadron continuity.
Recall that the existence of the anomaly-induced critical point opens up the possibility to go, at least at zero temperature, smoothly from the 
COE phase to the
highest-density phase, the (approximate) CFL phase. If we now introduce a meson-condensed phase we might introduce an additional phase 
transition, separating the COE from the COE-$K^0$ phase. This phase transition cannot end at a critical point because the kaon 
condensate breaks strangeness conservation $U(1)_S$ which is an exact symmetry of QCD, i.e., the CFL phases with and without kaon condensation have 
distinct residual symmetry groups. 
Only if we take into account the weak interaction which breaks flavor conservation, this line can end at a critical point. Another
way of saying this is that through weak interactions the Goldstone mode associated with kaon condensation receives a small mass which has 
been estimated to be of the order of $50\,{\rm  keV}$ \cite{Son:2001xd}. Here we do not consider such small $U(1)_S$-breaking terms and thus 
whenever both
COE and COE-$K^0$ phases are present in the phase diagram, they are separated by a true phase transition. Such an additional phase transition 
could be avoided if the COE phase is completely replaced by the COE-$K^0$ phase. 
\end{enumerate}

\subsection{Critical point(s) with meson condensate}

\begin{figure} [t]
\begin{center}
\hbox{\includegraphics[width=0.5\textwidth]{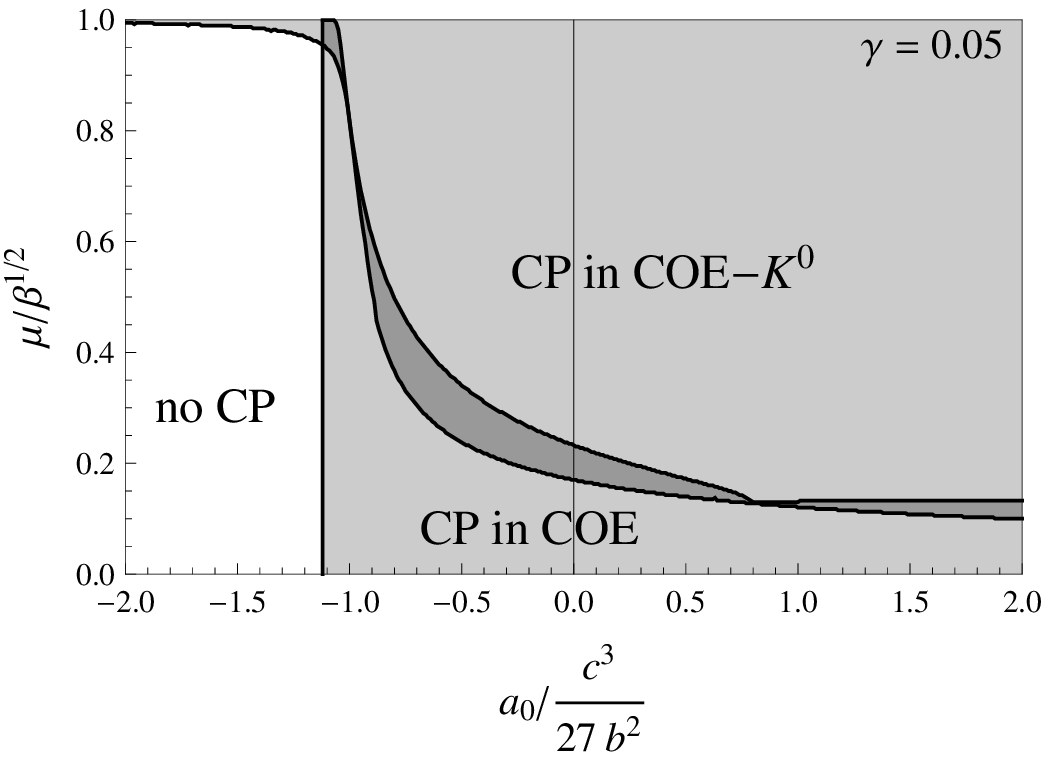}
\includegraphics[width=0.5\textwidth]{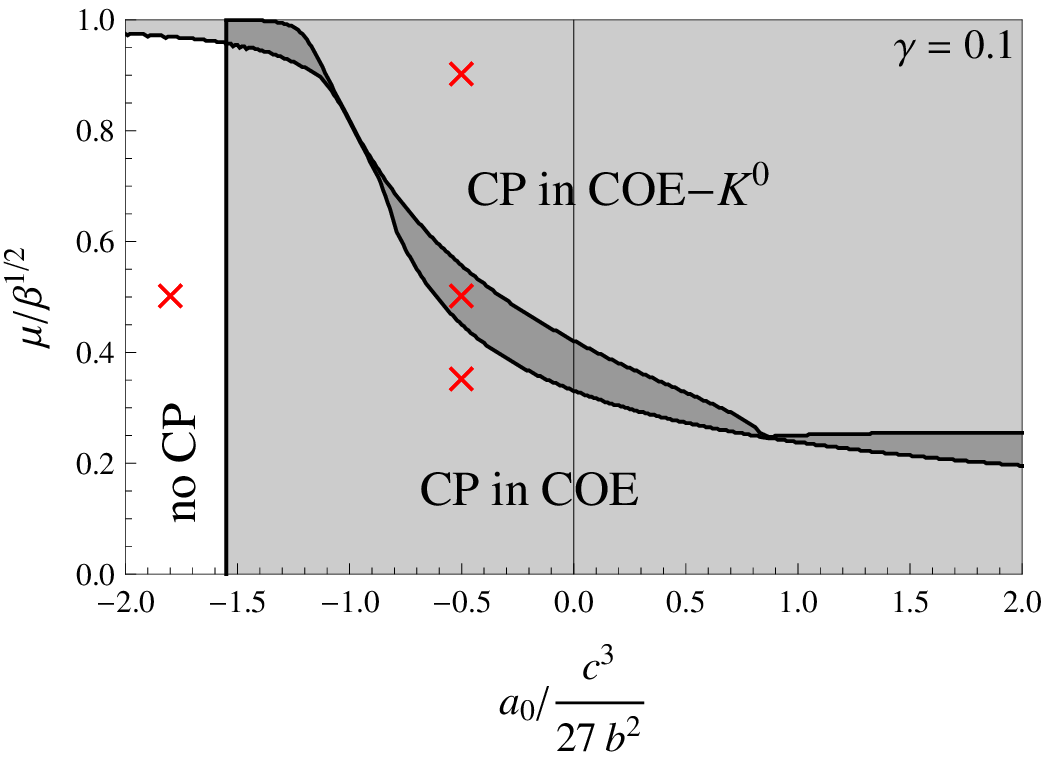}}
\caption{(Color online)
Classification of parameter regions in the $(a_0,\mu)$ plane for two different values of $\gamma$ according to the number of critical 
points (CP) in the $(a,\alpha)$ plane. Recall that $a_0$ is the coefficient of the linear term in the chiral condensate, induced by the 
strange quark 
mass, $\mu$ is the effective kaon chemical potential (bounded from above by $\mu=\beta^{1/2}$ beyond which our Ginzburg-Landau potential 
becomes unstable), and $\gamma$ parametrizes the strength of instanton effects. The plots show four qualitatively different 
cases each of which is represented by a diagram in Fig.\ \ref{figkaon}, corresponding to the four marked points in the right panel. 
In the dark grey regions there are, according to our numerical algorithm, two critical points. However, the corresponding phase diagrams show
that one of them seems to lie on top of the phase transition line between COE and COE-$K^0$, see upper right panel of Fig.\ \ref{figkaon}.}
\label{figmua0}
\end{center}
\end{figure}

In the following we shall always use fixed values for the parameters $b$, $c$, and $\beta$ with respect to which the results are insensitive. 
(We do not have a rigorous proof for that but we have made several numerical checks which suggest that variation of $b$, $c$, and $\beta$ does not 
change our main conclusions.) To answer question $(i)$
we thus determine for each value of the parameter set $(a_0,\mu,\gamma)$ the number of critical points in the $(a,\alpha)$ plane. To this end,
we compute the ground state at the two ``would-be'' critical points (\ref{K0crit}) and (\ref{COEcrit}) and check whether the COE-$K^0$
phase is the ground state at the point (\ref{K0crit}) and whether the COE phase is the ground state at the point (\ref{COEcrit}). 
The result is presented in Fig.\ \ref{figmua0} in the $(a_0,\mu)$ plane for two values of $\gamma$. 

The most obvious observation is
that for large $\mu$ it is more likely to find the critical point in the kaon-condensed phase. The reason is simply that with 
increasing $\mu$ the kaon-condensed phase covers more and more space in the phase diagram which supports our interpretation
of $\mu$ as an effective chemical potential. For sufficiently small values of 
$a_0$ there is no critical point at all (except for very large values of $\mu$, just below its maximum value, for which there is a critical 
point in the COE-$K^0$ phase). The reason is the same as already discussed for the COE phase in the lower right panel of Fig.\ \ref{figmass}:
both points (\ref{K0crit}) and (\ref{COEcrit}) are in a region where the NG phase is the ground state, and thus they are not realized. 
Fig.\ \ref{figmua0} shows that the region without critical points is shifted to smaller values of $a_0$ for increasing instanton effects,
parameterized by $\gamma$. We also see that with decreasing instanton effect, it becomes
more likely to find the critical point in the meson-condensed phase. 

The regions with a critical point in the COE phase are not separated from the regions with
a critical point in the COE-$K^0$ phase by a one-dimensional line. We rather find a two-dimensional region in the $(a_0,\mu)$ space
where our numerical algorithm finds two critical points, one in each phase. However, a closer look reveals that in this region 
(dark grey in Fig.\ \ref{figmua0}) the point (\ref{K0crit}) seems to lie exactly on top of the phase transition line between COE and COE-$K^0$. 
In other words, moving within the dark grey areas, the critical point seems to ``drag'' the second-order phase transition line which is attached 
to it. Fig.\ \ref{figmua0} also shows that this two-dimensional region is ``squeezed'' to a point in two instances. 
The first instance is at $\mu=(\frac{2}{3}\beta)^{1/2}$, $a_0=-\frac{c^3}{27b^2}$. From Eqs.\ (\ref{K0crit}) and (\ref{COEcrit}) we see that
for these parameter values the two potential critical points coincide and sit at $\alpha=0$, i.e., on the $a$ axis. On the other hand, 
Eq.\ (\ref{sigmacr}) shows that for $\mu=(\frac{2}{3}\beta)^{1/2}$ the phase transition line between COE and COE-$K^0$ is identical with the
$a$ axis because in this case $\sigma$ can only be finite if $\alpha=0$. Consequently, both points (\ref{K0crit}) and (\ref{COEcrit}) 
coincide and lie on the phase 
separation line. Now, keeping $\mu$ fixed and varying $a_0$ will keep both points together but move them away from the $a$ axis, while
the phase transition line remains unchanged. Hence, depending on which direction in $a_0$ one takes, a critical point will appear either 
in the COE (smaller $a_0$) or the COE-$K^0$ (larger $a_0$) phase. The second instance is at positive $a_0$, and the scenario is quite different 
here. Now the two points (\ref{K0crit}) and (\ref{COEcrit}) do not coincide. Nevertheless, for the given parameters (for $\gamma=0.1$ we read off 
$\mu\simeq 0.25\beta^{1/2}$, $a_0\simeq 0.85 \frac{c^3}{27b^2}$) both points lie on the second-order phase transition line between COE and 
COE-$K^0$, and again varying the parameters by an arbitrarily small amount creates a critical point in one or the other phase. In contrast to the 
first instance this is a purely numerical observation.

\begin{figure} [t]
\begin{center}
\hbox{\includegraphics[width=0.5\textwidth]{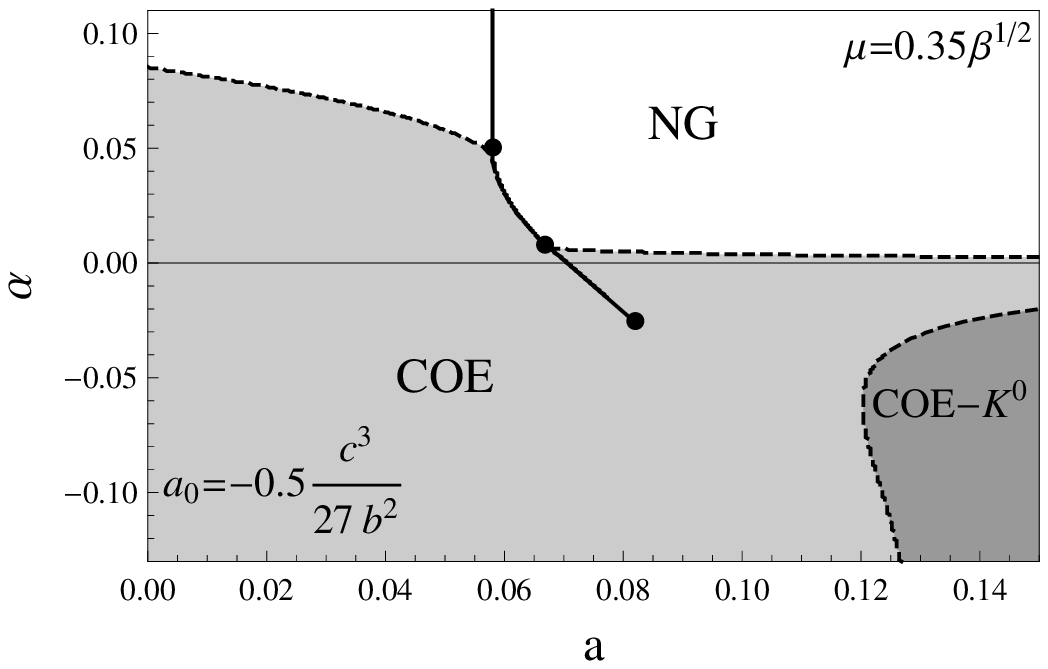}
\includegraphics[width=0.5\textwidth]{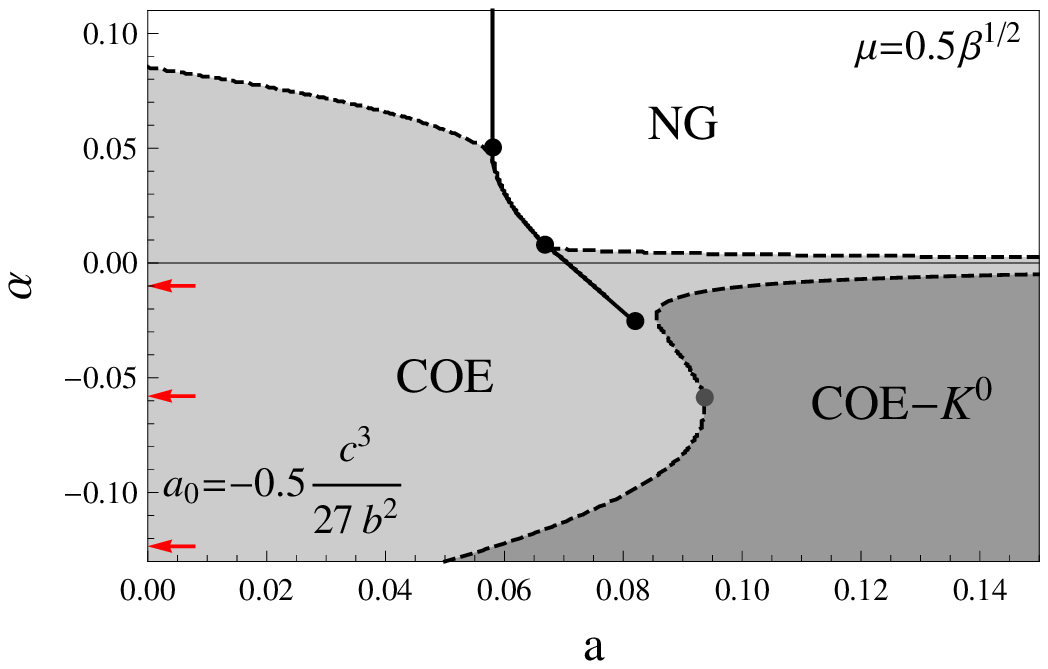}}

\hbox{\includegraphics[width=0.5\textwidth]{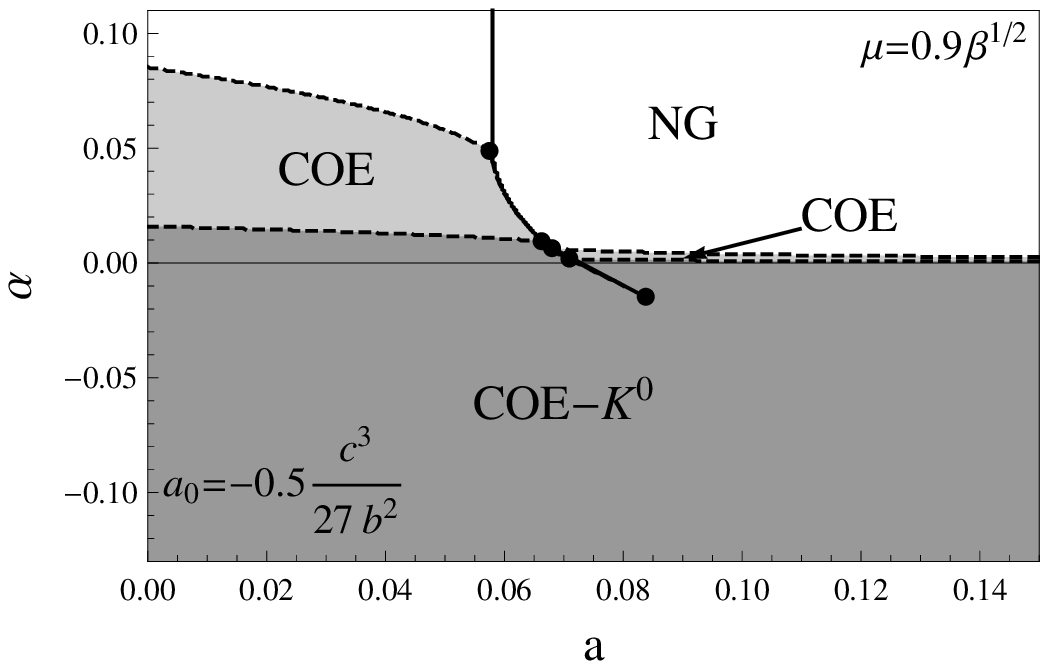}
\includegraphics[width=0.5\textwidth]{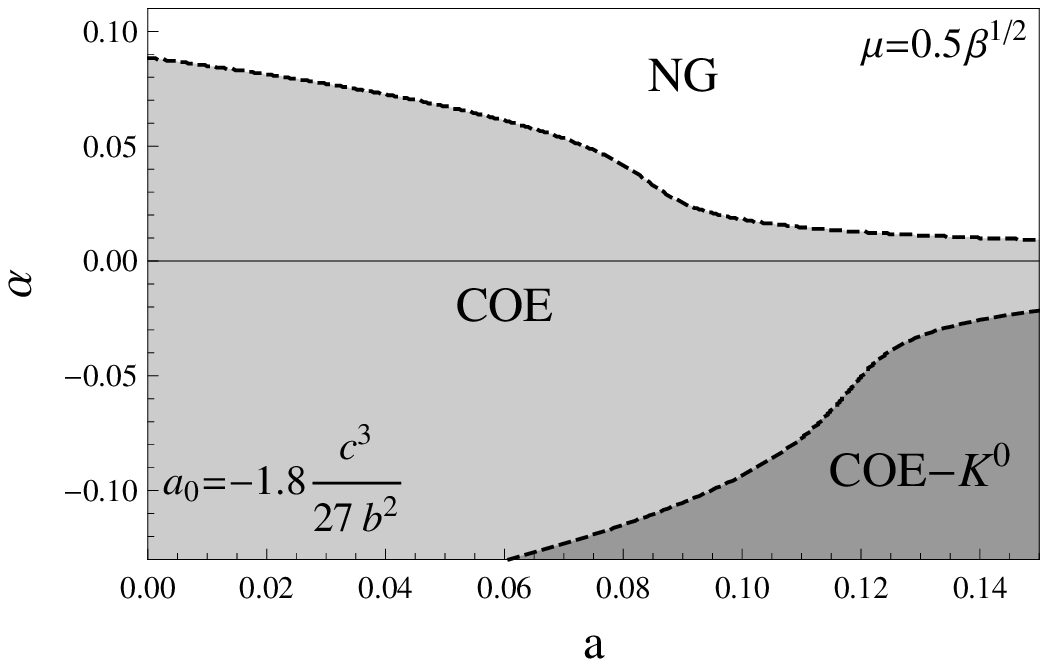}}
\caption{(Color online) 
Four Ginzburg-Landau phase diagrams including strange quark mass effects and kaon condensation, according to the potential (\ref{OmK0}). 
The parameters $b$, $c$, $\beta$ and
$\gamma$ are chosen as in Fig.\ \ref{figmass}, while $a_0$ and $\mu$ are given in each panel and correspond to the marked points in the 
right panel of Fig.\ \ref{figmua0}. The (red) arrows in the upper right panel indicate the $\alpha$ values for which the order
parameters are plotted in Fig.\ \ref{figsigmad} as functions of $a$. 
}
\label{figkaon}
\end{center}
\end{figure}

\subsection{Phase diagrams with meson condensate}

The four qualitatively different scenarios obtained from Fig.\ \ref{figmua0} are represented in the four panels of Fig.\ \ref{figkaon},
corresponding to the marked points in the right panel of Fig.\ \ref{figmua0}. We have checked that our phase transition lines between COE and 
COE-$K^0$ in Fig.\ \ref{figkaon}, which are obtained by a brute-force comparison of the free energies, are reproduced by the 
curves $\alpha(a)$ discussed below Eq.\ (\ref{sigmacr}). Together with the thin dashed lines in Fig.\ \ref{figmass} 
which reproduce the phase transition lines between the COE and NG phases, this means that we have a relatively simple, semi-analytical,
form for all second-order lines.
 
The first three panels (upper left \& right, lower left) show the phase diagram for increasing $\mu$ and all other parameters fixed. Not
surprisingly, the COE-$K^0$ phase covers more and more phase space with increasing $\mu$. More interestingly, for all values of $\mu$ that are 
allowed in our approximation, a finite region of the COE phase without meson condensation survives. Here we have increased $\mu$ up to 
90\% of its upper limit, but we have checked that this conclusion remains valid for all allowed values of $\mu$. This seems to answer the above
question $(ii)$ with no, and meson condensation always appears to induce an additional phase transition line which does not end at a critical 
point. We have checked, however, that this statement depends on our approximation. Taking into account additional terms, for instance
the ones proportional to $\alpha_2$, $\beta_2$, $\gamma_2$ in Eq.\ (\ref{Omfull}), it is possible to find regions in the parameter space where
the COE-$K^0$ phase completely eliminates the COE phase from the phase diagram. 

\begin{figure} [t]
\begin{center}
\hbox{\includegraphics[width=0.33\textwidth]{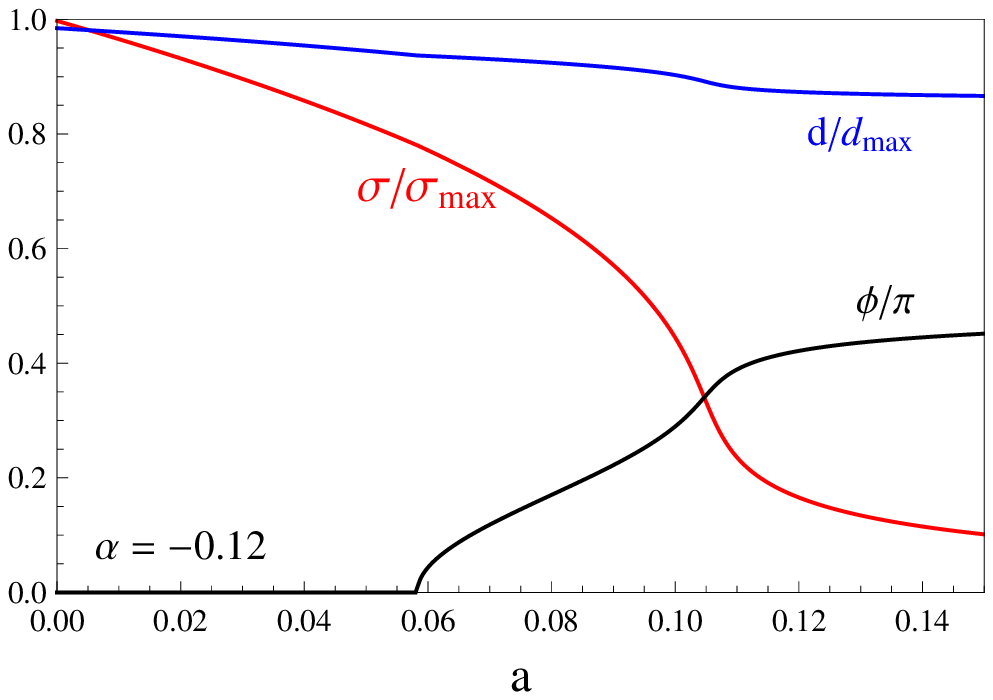}
\includegraphics[width=0.33\textwidth]{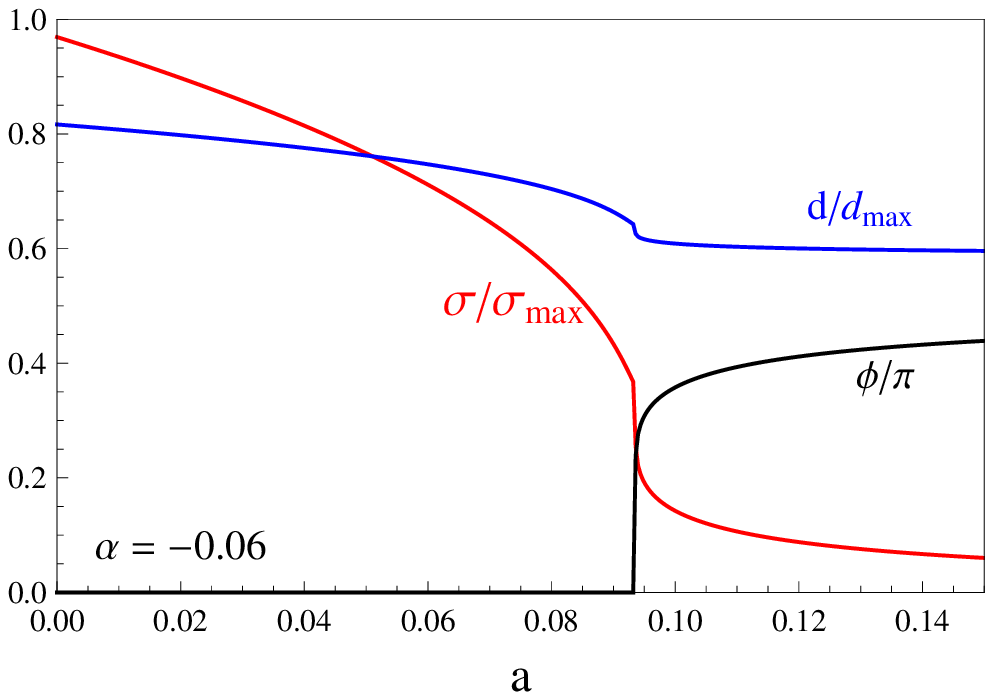}
\includegraphics[width=0.33\textwidth]{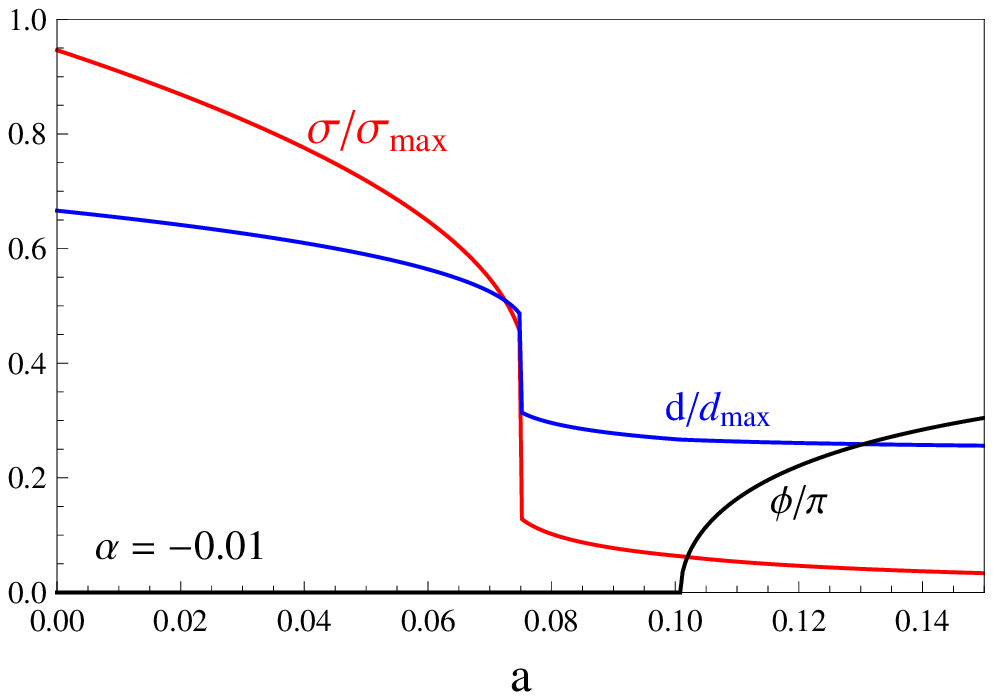}}
\caption{(Color online)
Order parameters $\sigma$, $d$, and $\phi$ for three values of $\alpha$ as functions of $a$, corresponding to the phase diagram in the 
upper right panel of Fig.\ \ref{figkaon}. As a normalization for the chiral and diquark condensates we have chosen their maximal values in the 
selected section of the $(a,\alpha)$ phase diagram. These values $\sigma_{\rm max}$ and $d_{\rm max}$ are assumed in the lower left corner, i.e., 
at $(a,\alpha)=(0,-0.13)$. The meson condensate is normalized by its maximum value $\pi$. Left (right) panel: onset of kaon condensation and
crossover (first-order phase transition) to a phase where the approximate $\mathbb{Z}_L(2)\times\mathbb{Z}_R(2)$ symmetry is restored happen at 
different values for $a$. Middle panel: these two transitions happen simultaneously.
}
\label{figsigmad}
\end{center}
\end{figure}

The lower left panel shows that for large $\mu$ the COE
phase survives in two disconnected regions. From the translation of the $(a,\alpha)$ plane into the QCD phase diagram, as discussed in 
Fig.\ \ref{figQCD}, we can expect the $T=0$ axis to pass through the larger of these two regions, on the left-hand side of the first-order
transition. The smaller strip on the right-hand side can be expected to be passed upon heating up the CFL phase, in agreement with 
NJL model calculations \cite{Warringa:2006dk}. In both regions it is interesting to check whether a less symmetric
color-superconducting phase than CFL becomes favorable. We discuss this possibility in the next section where we include the 2SC phase in 
our calculation.

The upper right panel is representative for all points in the dark grey area in Fig.\ \ref{figmua0}. Although our numerical algorithm shows
that the COE-$K^0$ phase is the ground state at the point (\ref{K0crit}), this appears not to be a critical point since there is no 
first-order line attached to it. Numerically we find that the second-order phase transition between COE and COE-$K^0$ in the vicinity of this
point is very strong, i.e., the kaon condensate $\phi$ develops a sizable nonzero value on a much smaller parameter region than it does
further away from this point. This can be seen in Fig.\ \ref{figsigmad} where we plot the order parameters $\sigma$, $d$, and $\phi$ as
a function of $a$ for three fixed values of $\alpha$, as indicated in the upper right panel of Fig.\ \ref{figkaon}. The middle panel of 
Fig.\ \ref{figsigmad} shows the behavior close to the point (\ref{K0crit}). We see that also the curves for $\sigma$ and $d$ are not
smooth around this point.

The curves for the order parameters also illustrate the first-order transition at large, but still negative, $\alpha$ (right panel) 
and its smooth version at small $\alpha$ (left panel). Translated to the QCD phase diagram, we can think of the latter, if present at all, 
as being closer to zero temperature as the former. 
In the case of the first-order transition, here taking place in the COE phase, both $\sigma$ and $d$ are affected significantly. After the 
transition, the chiral condensate goes to zero for large $a$. This is as expected because the phase at large $a$ corresponds to the 
(approximate) CFL-$K^0$ phase. 
For the crossover, here taking place in the CFL-$K^0$ phase, we see that the diquark and meson condensates are not much affected, only 
the chiral condensate decreases smoothly but drastically. The location of this crossover is given by the continuation of the critical line, 
see Fig.\ \ref{figCP}. 
 
We can rephrase the main conclusion from Fig.\ \ref{figsigmad} in the following concise way. There are basically two transitions: in the first, 
the chiral condensate
goes to approximately zero; this is either a first-order transition or a crossover since the symmetry which gets restored is only approximate in the 
presence of the axial anomaly. The second is the onset of kaon condensation which is always of second order since the broken symmetry
is exact (neglecting weak interactions). The transitions appear at two separate points in the left as well as in the right panel
(and in different orders, comparing left with right). In the middle panel, they appear approximately at the same value of $a$, which 
seems to be the reason for the interesting, non-smooth, behavior in this case.

\section{Under which conditions does the 2SC phase appear?}
\label{sec:2SC}

So far our choice of the color-superconducting phases was inspired by high-density arguments. We have considered the 
CFL phase, which is present at asymptotically large densities, and the kaon-condensed CFL phase, which is the first adjustment of the 
CFL phase to the effect of a small strange quark mass within a weak-coupling approach. Our calculation, however, intends to shed light on the 
phase structure at moderate densities where less symmetric phases may appear, as discussed in the introduction. In this section we take into
account one of these phases, namely the 2SC phase. In the 2SC phase, all strange quarks as well as all quarks of one color, say blue, 
remain unpaired, i.e., Cooper pairs are made of red up/green down and green up/red down quarks. At weak coupling and 
parametrically small strange quark mass the 2SC phase has larger free energy than either CFL or unpaired quark matter \cite{Alford:2002kj}.
Phenomenological models such as the NJL model suggest that this may no longer be true at large coupling \cite{Ruster:2005jc,Abuki:2005ms},
and the 2SC phase (or variants thereof) may cover a region in the phase diagram between the low-density chirally broken phase and CFL. 
The appearance of the 2SC phase would clearly interrupt a possible quark-hadron continuity since 2SC does not break chiral symmetry and thus
true phase transitions would be unavoidable between hadronic matter and 2SC and between 2SC and CFL. Building on NJL model calculations
with $U(1)_A$-breaking terms \cite{Abuki:2010jq,Abuki:2010ej}, it has been argued that the 2SC phase 
indeed covers the potential anomaly-induced critical point for a wide region in the NJL parameter space \cite{Basler:2010xy}\footnote{After
completing this work, we have learned that in a new version (v2) of Ref.\ \cite{Basler:2010xy} an appendix containing a discussion of the 
2SC phase in a Ginzburg-Landau approach, with some overlap to this section of our work, has been added.}.

To get an idea about the possibility of a 2SC phase in our general Ginzburg-Landau formalism, we discuss the 2SC phase in the simplest possible way. 
We shall not attempt to study the whole phase space with 2SC and meson-condensed CFL. This would require the use of several additional 
Ginzburg-Landau parameters. 
Without meson condensation we shall be able, however, to make some general statements about the
phase diagram including 2SC. On a qualitative level, we give some arguments about the addition of meson condensation
at the end of this section. We also do not attempt to account for electric and color neutrality and beta equilibrium. These conditions are crucial 
for the appearance of non-CFL color superconductors since they impose constraints on the Fermi momenta of the various quark species.
So far our phase diagrams have included only the CFL and CFL-$K^0$ phases, and are thus not expected to be affected much by these constraints 
because the symmetric CFL pairing pattern ensures, at least at $T=0$, that the number density of all quark species is the same. For the phase
diagrams including the 2SC phase, to be discussed in this section, the neutrality constraint is more important. In our general Ginzburg-Landau 
approach, however, electric (or any other) charge cannot be defined. This is only possible if an explicit dependence 
of the Ginzburg-Landau parameters on the chemical potentials is assumed \cite{Iida:2000ha}, for instance using results from perturbative QCD or 
a phenomenological model. Here we keep the parameters general and determine their range where the 2SC phase appears in the phase diagram. 
Neutrality and beta equilibrium can then be expected to yield relations between the parameters and thus define an ``allowed'' 
(neutral and beta-equilibrated) subspace of the full parameter space. 

Without meson condensate, the CFL order parameter is simply $d_L=d_R={\rm diag}(d,d,d)$, which is obtained from the 
more general order parameter (\ref{dLdR}) by setting $\phi=0$. For the 2SC phase, the order parameter is 
$d_L=d_R={\rm diag}(0,0,d)$ which describes pairing of only up and down quarks of two colors. In order to compare the free energies of 
2SC and CFL we need to go 
back to the general Ginzburg-Landau terms. We shall for simplicity keep our assumption $\Phi={\rm diag}(\sigma,\sigma,\sigma)$ in both 2SC and CFL.
As an alternative ansatz, accounting for the broken flavor symmetry, one might use the ansatz $\Phi={\rm diag}(\sigma,\sigma,0)$ for the 
2SC phase. In this case, the potential becomes trivial because the $d^2\sigma$ term that couples chiral and diquark condensates vanishes, 
and we have checked numerically that the 2SC phase appears nowhere in the phase diagram. 
For a more complete study one would have to include the $d^2\sigma^2$ 
interaction term and/or allow for independent chiral condensates $\Phi={\rm diag}(\sigma_u,\sigma_d,\sigma_s)$. Here we proceed with the 
symmetric ansatz for $\Phi$ and show that the 2SC phase appears in certain regions of the parameter space 
in accordance with physical expectations and with NJL studies. 

Within this ansatz, the chiral part
of the potential $\Omega_\Phi$ can be taken directly from Sec.\ \ref{sec:GL} and is the same for CFL and 2SC. With the 
help of Eqs.\ (\ref{dtraces}), (\ref{Mdd}), and (\ref{md4}), we write the general form of the diquark part as
\bea
\Omega_d &=& \alpha_1\left(\Tr[d_Ld_L^\dag]+\Tr[d_Rd_R^\dag]\right) + \alpha_2\left(\Tr[d_L^\dag Md_R]+{\rm h.c.}\right) \non
&& +\, \beta_1\left\{(\Tr[d_Ld_L^\dag])^2+(\Tr[d_Rd_R^\dag])^2\right\} +\beta_2\Tr[d_Ld_L^\dag]\Tr[d_Rd_R^\dag]
+\beta_3\left\{\Tr[(d_Ld_L^\dag)^2]+\Tr[(d_Rd_R^\dag)^2]\right\} \non
&&+\, \beta_4\Tr[d_Rd_L^\dag d_Ld_R^\dag]  
+\beta_5\left(\Tr[d_L^\dag Md_Rd_R^\dag d_R]+{\rm h.c.}\right) + \beta_6\left(\Tr[d_R^\dag Md_Ld_L^\dag d_L]+{\rm h.c.}\right)\non
&&+\,\beta_7\left(\Tr[d_L^\dag M d_R]\Tr[d_R^\dag d_R]+{\rm h.c.}\right)+\beta_8\left(\Tr[d_L^\dag M d_R]\Tr[d_L^\dag d_L]+{\rm h.c.}\right) \, ,
\eea
where we have assumed the coefficients in front of terms that are related by an exchange of $L$ and $R$ to be identical. Here, 
$\alpha_1$ and $\alpha_2$ are the coefficients for the $d^2$ terms without and with mass insertion, and $\beta_1,\ldots,\beta_4$ and
$\beta_5,\ldots,\beta_8$ are the coefficients for the $d^4$ terms without and with mass insertions.
For the $d^2\sigma$ interaction terms (we neglect again the $d^2\sigma^2$ terms) we obtain from Eqs.\ (\ref{ddPhi}) and 
(\ref{ddMPhi})
\bea
\Omega_{\Phi d} &=& \gamma_1\left(\Tr[d_Rd_L^\dag\Phi]+{\rm h.c.}\right) 
+ \gamma_2\left(\Tr[d_Ld_L^\dag+d_Rd_R^\dag]\Tr[M^\dag\Phi]+{\rm h.c.}\right) 
+\gamma_3\left(\Tr[d_Ld_L^\dag M\Phi]+{\rm h.c.}\right) \non
&&+\, \gamma_4 \left[\epsilon_{abc}\epsilon_{ijk}M_{ai}\Phi_{bj}(d_Ld_R^\dag)_{ck}+{\rm h.c.}\right] \, ,
\eea
with $\gamma_1$ and $\gamma_2,\ldots,\gamma_4$ being the coefficients for the terms without and with mass insertions. 
After performing the traces we can write the 2SC and CFL potentials as
\begin{figure} [t]
\begin{center}
\hbox{\includegraphics[width=0.5\textwidth]{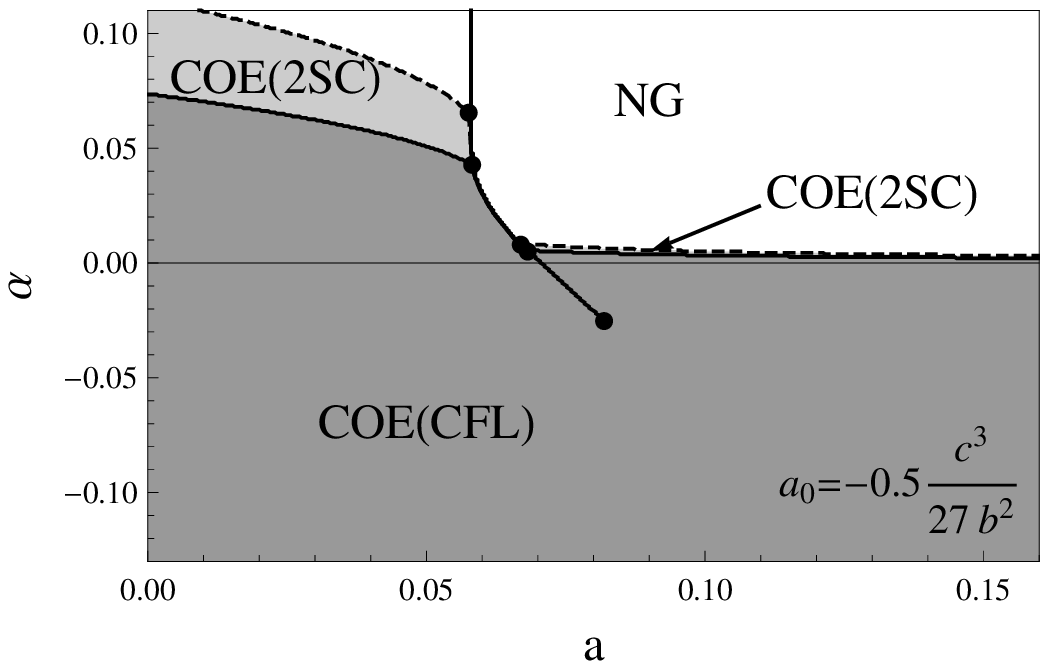}
\includegraphics[width=0.5\textwidth]{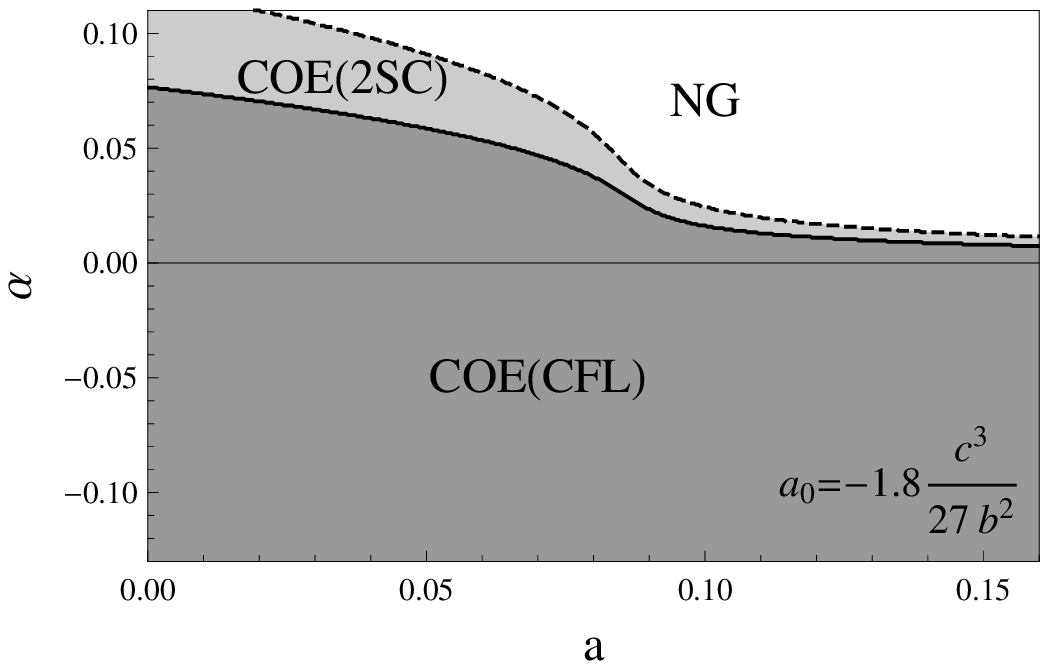}}
\caption{Ginzburg-Landau phase diagrams including the 2SC phase for two different values of $a_0$ 
(one should think of increasing the strange quark mass from left to right) and all other parameters fixed as in 
Fig.\ \ref{figmass}. The 2SC phase occurs after including mass 
corrections according to the potentials (\ref{CFL2SC1}), for these 
plots we have chosen $-\tilde{\alpha}m_s=\tilde{\gamma}m_s=0.05$. As a function of increasing $|a_0|$ ($a_0<0$) the 2SC phase first appears on the 
left-hand side of the phase transition between NG and COE, then additionally in a disconnected region on the right-hand side (left panel), before
the two regions merge for sufficiently large $|a_0|$ (right panel). COE(CFL) denotes the phase with nonzero chiral condensate 
and diquark
condensate in the CFL phase (denoted simply COE in all previous plots), COE(2SC) denotes the phase with nonzero chiral condensate and diquark
condensate in the 2SC phase. Had we not taken into account the 2SC phase, the (second-order) transition between COE and NG would have been 
between the shown COE(CFL)/COE(2SC) and COE(2SC)/NG transitions, see upper and lower right panels of Fig.\ \ref{figmass}. In this sense,  
the 2SC phase extends the color-superconducting area. 
}
\label{fig2SC}
\end{center}
\end{figure}
\begin{subequations}\label{CFL2SC}
\bea
\Omega_{{\rm CFL}} &=& \Omega_\Phi + (\alpha_1+\alpha_2 m_s) d^2 + (\beta_1^{\rm CFL}+\beta_2^{\rm CFL}m_s) d^4 
+ (\gamma_1+\gamma_2^{\rm CFL}m_s)d^2\sigma \, , \\
\Omega_{{\rm 2SC}} &=& \Omega_\Phi + \left(\frac{\alpha_1}{3}+\alpha_2 m_s\right) d^2 + (\beta_1^{\rm 2SC}+\beta_2^{\rm 2SC}m_s) d^4 
+ \left(\frac{\gamma_1}{3}+\gamma_2^{\rm 2SC}m_s\right)d^2\sigma\, ,
\eea
\end{subequations}
with $\Omega_\Phi$ given in Eq.\ (\ref{OmPhi}).
We see that for the $d^2$ and $d^2\sigma$ terms we know the ratio of the coefficients between 2SC and CFL for the mass-independent terms.
Without mass corrections we can thus simply use 1/3 of the coefficient of the CFL phase to obtain the corresponding term for the 
2SC phase. Including a small mass term then corresponds to a small correction to this 1/3. (In this spirit, it is irrelevant that we also know
the ratio between the coefficients of the $m_sd^2$ terms.) Such a statement is not possible for the $d^4$
terms where in general we need independent parameters even for the mass-independent terms. Therefore, the overall coefficient in front of the 
$d^4$ term in the 2SC phase must be taken as a new parameter and cannot be expressed in 
terms of a single coefficient of the CFL phase. As a result, we can write the free energies in the following convenient way,
\begin{subequations}\label{CFL2SC1}
\bea
\Omega_{{\rm CFL}}(\sigma,d) &=& 
a_0\sigma + \frac{a}{2}\sigma^2-\frac{c}{3}\sigma^3 +\frac{b}{4}\sigma^4 + \frac{\alpha}{2} d^2 + \frac{\beta}{4} d^4  -\gamma d^2\sigma \, ,\\ 
\Omega_{{\rm 2SC}}(\sigma,d) &=& 
a_0\sigma + \frac{a}{2}\sigma^2-\frac{c}{3}\sigma^3 +\frac{b}{4}\sigma^4 + \frac{\alpha}{2}\left(\frac{1}{3}+\tilde{\alpha}m_s\right) d^2 + 
\frac{\tilde{\beta}}{4} d^4  -\gamma \left(\frac{1}{3}+\tilde{\gamma}m_s\right) d^2\sigma \, ,
\eea
\end{subequations} 
where $\alpha\equiv\alpha_1+\alpha_2 m_s$, $\beta\equiv\beta_1^{\rm CFL}+\beta_2^{\rm CFL}m_s$, $\gamma\equiv\gamma_1+\gamma_2^{\rm CFL}m_s$
in order to reproduce the potential from Eq.\ (\ref{Om0}). The new coefficients $\tilde{\alpha}$, $\tilde{\beta}$, $\tilde{\gamma}$
depend on the parameters in Eq.\ (\ref{CFL2SC}) in an obvious, but irrelevant, way.

We can now proceed analogously to the previous sections and determine the ground state of the system. Now we need to compare the NG, 
COE(CFL), and COE(2SC) phases, where COE(CFL) is the phase with coexisting chiral condensate $\sigma$ and diquark condensate in the 
CFL pattern (this phase was simply termed COE in the previous sections), and COE(2SC) is the phase where $\sigma$ coexists with the diquark 
condensate in the 2SC pattern.  
The main question is under which conditions and where in the phase diagram the 2SC phase appears. The results
of our numerical evaluation are summarized in the following observations and in Fig.\ \ref{fig2SC}.

\begin{itemize}
\item Our conclusions are insensitive to the value of $\tilde{\beta}$. Note that $\tilde{\beta}$ is the only new coefficient in the 2SC potential
that enters to leading order. It could thus have potentially complicated our results. We have checked, however, that variation of $\tilde{\beta}$ 
does not change any of the following statements and the topology of the phase diagrams shown in Fig.\ \ref{fig2SC}. For this
figure, we have thus simply chosen $\tilde{\beta}=\beta$.

\item If we neglect the mass corrections to the $d^2$ and $d^2\sigma$ terms, i.e., $\tilde{\alpha}=\tilde{\gamma}=0$, there is no 2SC phase
in our $(a,\alpha)$ phase diagram for all values of $a_0$. In other words, the mass effect through the linear term in the chiral potential
$\propto a_0$ is not sufficient to trigger the 
2SC phase. More mass corrections are needed.

\item  As soon as mass corrections $\tilde{\alpha}<0$ or $\tilde{\gamma}>0$ or both are switched on, there is at least one region in the 
$(a,\alpha)$ phase diagram for all $a_0$ where the 2SC phase is the ground state. (It is a numerical observation that only the given signs
of $\tilde{\alpha}$, $\tilde{\gamma}$ yield the results shown in Fig.\ \ref{fig2SC}; different signs, i.e., $\tilde{\alpha}>0$ and/or 
$\tilde{\gamma}<0$ require sufficiently large positive values of $a_0$ for the 2SC phase to appear.) 
The 2SC phase appears in the expected regions
of the phase diagram, separating the NG phase from the CFL phase. This is shown in Fig.\ \ref{fig2SC} for two different values of $a_0<0$. 
Increasing $|a_0|$ for negative $a_0$ increases the area which is covered by the 2SC phase. 

\item The phase transition from COE(CFL) to COE(2SC) is of first order. This is clear since two of the gap parameters must change 
discontinuously in the transition from $d_L=d_R={\rm diag}(d,d,d)$ to $d_L=d_R={\rm diag}(0,0,d)$. Additionally, our numerical 
results show that the chiral condensate is discontinuous at this transition. 

\end{itemize}

It is an interesting question whether meson condensation can prevent the 2SC phase from appearing. Comparing Figs.\ \ref{figkaon} and 
\ref{fig2SC}, the answer seems to be no because the COE-$K^0$ phase does not reach areas close to the transition to the NG phase, and this is
exactly where the 2SC phase lives. However, we have to remember that in our discussion of kaon condensation we have neglected all mass terms 
except for
$a_0$ and the one associated with the kaon chemical potential. The omitted terms are exactly the ones that are needed to obtain the 
2SC phase. Hence we would have to redo our analysis, including 2SC and kaon condensation and taking into account the mass corrections 
for the $d^2$ and $d^2\sigma$ terms in the potential (\ref{Omfull}). This would introduce more independent parameters and without 
further constraints a systematic study would be very unwieldy. Therefore, we have only done some numerical calculations with selected parameters. 
These calculations show that if we take for instance one of the phase diagrams 
in Fig.\ \ref{fig2SC}, there is a range of parameters for the mass corrections where the kaon-condensed phase does expel the 2SC phase.
However, the 2SC phase is only expelled completely from the $(a,\alpha)$ phase diagram if the mass corrections 
$\propto \alpha_2, \beta_2,\gamma_1,\gamma_2$ in Eq.\ (\ref{Omfull}) are of the order of or larger than the ${\cal O}(m_s^0)$ terms.
In this case, our Ginzburg-Landau expansion becomes unreliable since we have neglected mass terms of higher order (except for
the term $\propto \mu^2$ which is suggested to be relevant from high-density arguments, as explained). 
If we keep the mass corrections much smaller than the ${\cal O}(m_s^0)$ terms, the 2SC phase, if it is preferred over the CFL phase 
in some region of the phase diagram, also appears to be favored (in a smaller region) over the CFL-$K^0$ phase.

\section{Summary and outlook}
\label{sec:conclusions}

We have studied phases of dense matter in a Ginzburg-Landau approach. Previous Ginzburg-Landau studies have shown that the axial anomaly
may induce a high-density critical point in the QCD phase diagram, possibly leading to a smooth crossover between hadronic matter and color-flavor
locked quark matter. We have explained in detail that the existence of this critical point is a consequence of the (discrete) symmetry of the 
CFL phase which -- in the presence of the axial anomaly -- is not changed by adding a chiral condensate. Our main goal has been to 
extend the previous studies by including a strange quark mass. We have discussed several different, although related, mass effects. 

Firstly, the strange quark mass introduces a term linear in the 
chiral condensate $\sigma$, say $a_0\sigma$, which allows for a smooth crossover between the phases of broken and (approximately) 
restored chiral symmetry. This effect
is most relevant for the high-temperature, low-density phase of QCD where there is indeed such a crossover between the hadronic phase
and the Quark-Gluon Plasma, as we know from lattice calculations. We have shown that the term $a_0\sigma$ is also 
relevant for the high-density 
critical point. The reason is the anomalous interaction term that couples the chiral to the diquark condensate $d$, say $\gamma d^2\sigma$. 
For sufficiently large values of $|a_0|$ ($a_0<0$) the first-order phase transition line which, for nonzero $\gamma$, 
ends at the high-density critical point disappears. As a result the transition between the ordinary chirally broken phase and the CFL phase 
is smooth everywhere.

Secondly, a nonzero strange quark mass is expected to induce less symmetric color-superconducting phases. 
In high-density calculations, the CFL-$K^0$ phase is the first phase that appears after going down in density from the asymptotically 
dense CFL region. We have introduced a kaon condensate as a relative rotation of left- and right-handed diquark condensates and have adjusted 
the Ginzburg-Landau potential to match the essential terms of the high-density effective theory. We have identified the region in the 
parameter 
space where the critical point has moved from the CFL into the CFL-$K^0$ phase and have determined the location of both possible critical points
in the presence of a strange quark mass. In addition to a shift of the critical point, the kaon condensate introduces a true phase transition
because it breaks strangeness conservation spontaneously which is an exact symmetry in QCD. 

Thirdly, we have discussed a more radical reaction of the system to a nonzero strange quark mass, namely the appearance of the 2SC phase. 
In previous studies in the Ginzburg-Landau approach, it has been shown that for infinitely large strange quark mass there cannot be a 
high-density critical point. The reason is that the 2SC phase does not break chiral symmetry and thus there must be a true phase transition between 
the 2SC phase with and without coexisting chiral condensate. Since we have included a nonzero, but finite, strange quark mass, we could study 
the competition between the 2SC and CFL phases under the influence of a nonzero chiral condensate. We have shown
that the mass term $a_0\sigma$ is not sufficient to favor the 2SC phase in any part of the phase diagram. Additional
mass terms, which we have neglected in our discussion of the meson condensate, are necessary for the 2SC phase to appear between unpaired 
quark matter and the CFL phase. In a conjectured translation to the QCD phase diagram it seems that the 2SC phase appears ``first''
(i.e., for the smallest values of these mass terms) at low temperature. As a consequence, the smooth crossover at zero temperature between
hadronic and quark matter would be disrupted by true phase transitions. 
The appearance of the 2SC phase is in agreement with recent NJL model
calculations.

There are several possible extensions of our work. We have studied the competition between 2SC and CFL systematically, but have only briefly 
discussed the competition between 2SC and CFL-$K^0$. For a more complete analysis it would be helpful to first find some constraints for the 
additional Ginzburg-Landau parameters. 
The potential proliferation of parameters has also led us to a simplified, flavor-symmetric ansatz for the chiral condensate. 
One should check in further studies how our results change with a more realistic ansatz. 
It would also be interesting to consider different meson condensates and possibly their coexistence. This is important  
since we do not know the masses of the CFL mesons at intermediate densities, and it may well be a different meson than the kaon which 
condenses in this regime. Furthermore, one should also take into account the requirement of
electric and color neutrality. This was not an issue in our calculations with the CFL phase since this phase is 
automatically neutral (and a neutral kaon condensate does not change this). In the 2SC phase, however, the numbers of up, down, and strange 
quarks are not identical, and the existence and details of this phase (as for any non-CFL color superconductor) depend strongly on the neutrality 
constraint. 

More generally speaking, the model-independent Ginzburg-Landau approach, including possible extensions in the future, is helpful to gain 
insight 
into the QCD phase diagram at low temperature and large, but not asymptotically large, densities. Like NJL model calculations, however, it is 
far from being conclusive for the actual, full QCD situation. Therefore, it is important to 
also pursue other approaches such as improvements of perturbative calculations \cite{Kurkela:2009gj} or studies of dense matter in the astrophysical 
context and comparing properties of phases of dense (quark) matter with data from compact stars \cite{Ozel:2010fw,Steiner:2010fz,Kurkela:2010yk}.

\begin{acknowledgments}
The authors acknowledge valuable discussions with M.\ Alford, K.\ Rajagopal, T.\ Sch{\"a}fer, Q.\ Wang, and N.\ Yamamoto.
This work was initiated at the workshop "New Frontiers in QCD 2010", at the Yukawa Institute of Theoretical Physics, Kyoto, Japan.
The work of M.T. is supported in part by
Saga University Dean's Grant 2010 for
Promising Excellent Research Projects.
\end{acknowledgments}

\bibliography{refs}

\end{document}